\documentclass{article}
\usepackage{graphicx}
\usepackage{subfigure}
\usepackage{amsfonts}
\usepackage{amsmath}
\usepackage{amssymb}
\usepackage{url}
\usepackage{fancyhdr}
\usepackage{indentfirst}
\usepackage{enumerate}
\IfFileExists{dsfont.sty}{\usepackage{dsfont}}{}
\usepackage[colorlinks=true,citecolor=blue]{hyperref}
\usepackage{amsthm}
\usepackage{color}
\usepackage{pgfplots}
\pgfplotsset{compat=1.18}
\DeclareRobustCommand{\rev}[1]{{\color{blue}#1}}

\usepackage{natbib}
\usepackage{comment}
\usepackage{capt-of}

\def\d{\mathrm{d}}

\def\lawis{\buildrel \mathrm{d} \over \sim}

\newcommand{\D}{\mathcal {D}}

\newcommand{\qbf}{\mathbf{q}}
\newcommand{\pbf}{\mathbf{p}}
\newcommand{\bm}{\boldsymbol}

\newcommand{\E}{\mathbb{E}}

\newcommand{\R}{\mathbb{R}}
\newcommand{\N}{\mathbb{N}}
\newcommand{\p}{\mathbb{P}}

\newcommand{\M}{\mathcal{M}}

\newcommand{\ssd}{\preceq_{\mathrm{icx}}}

\renewcommand{\(}{\left(}
\renewcommand{\)}{\right)}

\renewcommand{\ge}{\geqslant}
\renewcommand{\le}{\leqslant}
\renewcommand{\geq}{\geqslant}
\renewcommand{\leq}{\leqslant}
\renewcommand{\epsilon}{\varepsilon}

\theoremstyle{plain}
\newtheorem{theorem}{Theorem}
\newtheorem{corollary}{Corollary}

\newtheorem{proposition}{Proposition}
\theoremstyle{definition}
\newtheorem{definition}{Definition}

\newtheorem{assumption}{Assumption}
\newtheorem{example}{Example}

\theoremstyle{remark}
\newtheorem{remark}{Remark}

\newcommand{\cet}{\begin{center}}
\newcommand{\ecet}{\end{center}}

\usepackage{setspace}
\newcommand{\com}[1]{\marginpar{{\begin{minipage}{0.18\textwidth}{\setstretch{1.1} \begin{flushleft} \footnotesize \color{red}{#1} \end{flushleft} }\end{minipage}}}}

\begin{document}

\title{Equity and Aggregate Ambiguity}%

\author{
Jose Blanchet\thanks{\scriptsize Department of Management Science and Engineering, Stanford University, USA. Email: \texttt{jose.blanchet@stanford.edu}}
\and
Yang Liu\thanks{\scriptsize 
School of Science and Engineering, The Chinese University of Hong Kong (Shenzhen)
, China. Email: \texttt{yangliu16@cuhk.edu.cn}} 
\and Tiantian Mao\thanks{\scriptsize Department of Statistics and Finance,
University of Science and Technology of China, China. Email: \texttt{tmao@ustc.edu.cn}}
\and Ruodu Wang\thanks{\scriptsize Department of Statistics and Actuarial Science, University of Waterloo, Canada. Email: \texttt{wang@uwaterloo.ca}}
}
\date{}
\maketitle


\begin{abstract}


Equitable resource allocations in the context of 
symmetric 
risk exposures are intuitively desirable and sometimes even expected in public planning. However, public planners generally are also concerned with mitigating risk and, more generally, uncertainty and ambiguity. 
We study a fundamental conflict between equity and ambiguity avoidance in public risk allocation problems. Our main results are framed and established in the setting of general symmetric aggregate losses equipped with dependence uncertainty among individual losses. Specializing to the classical model of \cite{K80}, we prove that an increase in equity is equivalent to an increase in aggregate ambiguity. We then provide a quantitative characterization of equity and ambiguity avoidance through the $\alpha$-maxmin preference framework, establishing conditions on ambiguity and risk attitudes under which each criterion holds. Specifically, equity corresponds to ambiguity or risk proneness, while ambiguity avoidance corresponds to ambiguity or risk aversion. The analysis extends to general classes of symmetric aggregation functions, where we show that ambiguity attitude is the primary determinant of preference, though aggregation structure and risk attitude also play important roles. Numerical illustrations validate our theoretical findings and explore the equity-ambiguity tradeoff. 

\end{abstract}



\section{Introduction}


Public risk allocation is a fundamental challenge for decision makers facing catastrophic events. As a motivating example, consider an earthquake threatening a region with $n$ buildings. A central planner has a fixed budget of protective resources to allocate among them. Each building $i$ faces a Bernoulli collapse risk $\tilde{x}_i \sim B_{1-\theta_i}$, where $\theta_i \in [0,1]$ represents the survival probability (or equivalently, the fraction of resources allocated to building $i$). The aggregate loss is the total number of collapsed structures and is denoted by $\tilde{d} = \sum_{i=1}^n \tilde{x}_i$. The distribution of $\tilde d$ depends on both the marginal survival probabilities (captured by the resource allocation profile $\boldsymbol{\theta} = (\theta_1, \ldots, \theta_n)$ with total budget $\bar{\theta} = \sum_{i=1}^n \theta_i$) and the dependence structure among building failures. The latter arises from geological factors such as epicenter location, soil composition, and building alignment, which may be  difficult to estimate in practice. To illustrate the key tension, consider two allocation policies: an inequitable profile $\boldsymbol{\gamma} = (1, 1, 0)$ concentrating resources on two buildings, and an equitable profile $\boldsymbol{\theta} = (2/3, 2/3, 2/3)$ distributing resources uniformly.  In this example, the planner who prefers equity, and thus favors $\boldsymbol{\theta}$ over $\boldsymbol{\gamma}$, must necessarily accept greater uncertainty about the aggregate outcome, which is random with an unknown distribution for $\boldsymbol{\theta}$, but deterministic and known for $\boldsymbol{\gamma}$. 

This motivating problem reflects several distinct but complementary real-world contexts. Additional examples include: 
\begin{itemize}
\item 
\textbf{Pandemic risk allocation:} A government distributes medical resources (vaccines, treatments, hospital capacity) across $n$ populations to reduce mortality risk, where the dependence of infection rates across regions depends on unknown transmission patterns and human mobility. 

\item 
\textbf{Environmental hazard planning:} Regulators allocate flood defenses or climate adaptation resources across urban and rural zones with heterogeneous vulnerability, where flood correlation across regions is uncertain due to complex hydrological dependence. 
\end{itemize}
In each setting, the planner must navigate a fundamental tension between two objectives: favoring \emph{risk equity} (equalizing protection across populations) and avoiding \emph{aggregate uncertainty} (minimizing uncertainty in total losses when dependence is unknown).

The classical framework for this problem originates with \cite{K80}, who studied two decision criteria. \emph{Catastrophe aversion} means the planner prefers a less dispersed loss distribution over a more dispersed one, holding the mean fixed. \emph{Risk equity} means the planner prefers allocations where survival probabilities are more ``balanced'' across individuals. That is, under the majorization order, individual risks are more similar given a fixed total budget. Under the assumption of \emph{independent} individual risks, \cite{K80} proved that catastrophe aversion and risk equity are fundamentally incompatible: any policy achieving greater equity necessarily increases the aggregate risk. \cite{F84} extended this analysis by proposing equity axioms for public risks and exploring different utility specifications. Later, \cite{BRT18} relaxed the independence assumption and showed that the conflict may disappear when individual losses exhibit interdependence, opening the door to policies simultaneously achieving both equity and catastrophe aversion.

Historically, public risk evaluation has been conducted within the framework of pure risk based on expected utility theory, following the classic line of  work of \cite{NM44}. A pure risk assumes that losses are random variables with known distributions. A critical dimension has been largely absent from this literature: \emph{ambiguity} (or \emph{Knightian uncertainty}), meaning that the exact distribution is unavailable to the decision maker. While ambiguity has been extensively studied in economics (\citealp{K21,E61,GS89,HS01,KMM05,MMR06}), its role in public risk allocation remains underexplored.

This paper addresses this gap by introducing ambiguity into the equity-catastrophe aversion framework. 
Within such a framework, ambiguity is naturally described in dependence. 
Rather than assuming a known dependence structure among individual losses, we consider the case where this dependence is fundamentally uncertain. This generates \emph{aggregate ambiguity} (uncertainty in the distribution of total losses) which becomes the focal point of analysis in place of the aggregate risk studied by \cite{K80}. Our main theoretical contributions are as follows. In Section \ref{sec:model}, we formalize a general framework where the social planner faces aggregate ambiguity. In Section \ref{sec:main_results}, we establish a striking conflict between equity and aggregate ambiguity avoidance (Theorems \ref{th:1}-\ref{th:2}) in this general setting. In Section \ref{sec:keeney-model}, we specialize to the classic \cite{K80} model and prove a stronger result: an increase in equity is \emph{equivalent} to an increase in aggregate ambiguity (Theorems \ref{thm:equiv}-\ref{thm:equiv2}). This means that a social planner who values equality among individuals necessarily faces greater uncertainty about the total loss when dependence is unknown. 
We show that this fundamental conflict holds universally across all aggregation functions, extending far beyond the sum-based framework of prior work. The technical foundation rests on recently developed tools of robust risk aggregation from \cite{CLLW22} and \cite{BLLW25}.
 As a consequence, equity and aggregate ambiguity avoidance emerge as fundamentally
incompatible objectives   in public risk evaluation. 
 We also clarify the relationship to classical results by showing that both increase in equity and increase in ambiguity imply increase in aggregate risk under independence. The converse does not hold (Proposition \ref{prop:2}), demonstrating that our uncertainty-based framework genuinely extends the equity-catastrophe aversion conflict of \cite{K80} and \cite{BRT18}.


To provide quantitative traction on these abstract concepts, we adopt the $\alpha$-maxmin preference framework (\citealp{GMM04}), which offers significant analytical advantages. In Section \ref{sec:preference}, we use this framework to characterize the social planner's decision criteria in terms of fundamental preference properties. Specifically, we show that equity is equivalent to either ambiguity proneness or risk proneness, while aggregate ambiguity avoidance is equivalent to either ambiguity aversion or risk aversion (Theorem \ref{thm:ambi}). 
Section \ref{sec:numerical} provides numerical illustrations of these theoretical findings. We further extend our analysis in Section \ref{sec:general} by exploring how different classes of aggregation functions (submodular, supermodular, and other symmetric forms) affect the equity-ambiguity tradeoff. The main insight is that ambiguity attitude is the essential determinant of preference, while aggregation structure and risk attitude introduce refinements to this fundamental distinction. 
Proofs and additional technical results are collected in Appendix \ref{sec:app_proof}.

\section{General model}
\label{sec:model}

\subsection{Allocation problem}

Let $\Theta \subseteq \mathbb{R}$ be the set of possible resource levels allocated to an individual, and $n \geq 2$ be a fixed integer representing the number of individuals. The  main task is to distribute a total amount of resources $\bar{\theta} > 0$ among $n$ individuals. Write $[n] = \{1, \ldots, n\}$. Throughout, we use the notation $\boldsymbol{\theta} = (\theta_1, \ldots, \theta_n) \in \Theta^n$, $\boldsymbol{\gamma} = (\gamma_1, \ldots, \gamma_n) \in \Theta^n$, and $\mathbf{x} = (x_1, \ldots, x_n) \in \mathbb{R}^n$ without further specification. A feasible \emph{allocation of resources} is a vector
$$\boldsymbol{\theta} = (\theta_1, \ldots, \theta_n) \in \Theta^n \quad \text{such that} \quad \sum_{i=1}^n \theta_i = \bar{\theta}.$$
We call $\boldsymbol{\theta}$ a \emph{resource allocation profile}. The feasible set is denoted by
$$\mathcal{A}(\bar{\theta}) := \left\{\boldsymbol{\theta} \in \Theta^n : \sum_{i=1}^n \theta_i = \bar{\theta}\right\}.$$
All choices by the social planner are made among vectors in $\Theta^n$ with fixed sum (resource allocation profiles); they  are consequences of the latent preferences over   random variables, distributions,  and ambiguity sets.

Given an allocation $\boldsymbol{\theta} \in \mathcal{A}(\bar{\theta})$, individual $i$ incurs a loss $\tilde{x}_i$ with distribution function $F_{\theta_i}$. The aggregate loss to society is given by the random variable $H(\tilde{x}_1, \ldots, \tilde{x}_n)$, where $H : \mathbb{R}^n \to \mathbb{R}$ is a Borel-measurable, increasing and symmetric aggregation function. A function $H$ is \emph{symmetric} if $H(\mathbf{x}) = H(\pi(\mathbf{x}))$ for any $\mathbf{x} \in \mathbb{R}^n$ and any permutation $\pi$ on $[n]$. The symmetry assumption ensures that each individual is equally weighted in the aggregation; the sum is a natural example.

Since the dependence structure among $(\tilde{x}_1, \ldots, \tilde{x}_n)$ is unknown, the planner faces ambiguity (also called uncertainty in this paper) about the distribution of the aggregate loss. This ambiguity is captured by the \emph{aggregation set}
\begin{equation}\label{eq:aggregation-new}
\mathcal{D}_H(\theta_1, \ldots, \theta_n) = \left\{\text{cdf of } H(\tilde{x}_1, \ldots, \tilde{x}_n) : \tilde{x}_i \sim F_{\theta_i}, \; i = 1, \ldots, n\right\},
\end{equation}
where the set is taken over all possible dependence structures among the individual losses. The set $\mathcal{D}_H(\boldsymbol{\theta})$ thus represents the \emph{aggregate ambiguity} faced by the planner under allocation $\boldsymbol{\theta}$.

\subsection{Equity}

Since each individual is treated symmetrically by the planner, an allocation is considered more equitable if resources are distributed more evenly across the population. We formalize this through the majorization order, which generalizes the notion of equity studied in prior work on public risk allocation.

\begin{definition}[Equity]\label{def:resource-equity}
Let $\boldsymbol{\theta}, \boldsymbol{\gamma} \in \mathbb{R}^n$ represent two resource allocations. We say that $\boldsymbol{\theta}$ is \emph{more equitable} than $\boldsymbol{\gamma}$ if 
$$\sum_{i=1}^n \phi(\theta_i) \leq \sum_{i=1}^n \phi(\gamma_i) \quad \text{for any convex function } \phi;$$
equivalently, $\boldsymbol{\theta}$ is dominated by $\boldsymbol{\gamma}$ in majorization order, denoted
$\boldsymbol{\theta} \leq_m \boldsymbol{\gamma}$.
\end{definition}

The majorization relation $\boldsymbol{\theta} \leq_m \boldsymbol{\gamma}$ implies that both allocations have the same total resources: both lie in $\mathcal{A}(\bar{\theta})$ for some $\bar{\theta}$. Intuitively, $\boldsymbol{\theta} \leq_m \boldsymbol{\gamma}$ means that resources are more evenly dispersed under $\boldsymbol{\theta}$ than under $\boldsymbol{\gamma}$, holding the total constant. This equity concept aligns with standard usage in welfare economics; see \cite{K80} and \cite{BRT18}.

\begin{remark}
The majorization order is equivalent to a sequence of Pigou--Dalton transfers; see \cite{K80} for the foundational treatment. Formally, a Pigou--Dalton transfer moves a small amount of resources from a better-endowed individual to a worse-endowed one, without reversing their ranking or changing the total. Specifically, given $\boldsymbol{\gamma}$ with $\gamma_i < \gamma_j$, a Pigou--Dalton transfer produces $\boldsymbol{\theta}$ where
\begin{itemize}
\item [(i)] $\theta_i = \gamma_i + \delta$ and $\theta_j = \gamma_j - \delta$ for some $\delta \in (0, (\gamma_j - \gamma_i)/2]$; and
\item [(ii)]all other components remain unchanged.
\end{itemize}
By standard majorization theory (e.g., Section 1.A.3 of \citealp{MOA11}), $\boldsymbol{\theta}$ is more equitable than $\boldsymbol{\gamma}$ if and only if $\boldsymbol{\theta}$ can be obtained from $\boldsymbol{\gamma}$ by a finite sequence of Pigou--Dalton transfers and permutations. 
The characterization through Pigou--Dalton transfers is intuitive to explain why the majorization order underpins  the concept of equity.
\end{remark}

\section{Main results}

\label{sec:main_results}

\subsection{Aggregate ambiguity}

The set $\D_H(\bm \theta)$ defined by \eqref{eq:aggregation-new} represents the planner's ambiguity about the distribution of the
aggregate loss under allocation $\bm \theta$. A larger set $\D_H(\bm \theta)$ can be interpreted as
more ambiguity about the aggregate outcome: there are more distinct distributions that may
arise from different dependence structures among the individual losses.

Let $\mathcal M(\R)$ be the set of all distributions (represented by cdfs) on $\R$.
To compare two elements  $F_G, F_B$ in $\mathcal M(\R)$, a common tool is the first-order stochastic dominance of $F_B$ (bad) over $F_G$ (good), denoted by $F_B \ge_{\rm st} F_G$, which is defined as $F_B(x) \leq F_G(x)$ for any $x \in \R$. For $\tilde x\lawis F_G$ and $\tilde y\lawis F_B$, the relation $F_B \ge_{\rm st}F_G$ means
$\E[\ell(\tilde y)] \ge \E[\ell(\tilde x)]$ for any increasing function $\ell$, and hence $\tilde y$ is worse than $\tilde x$ for any decision maker with an increasing loss function.

A subset of $ \mathcal M(\R)$
describes all possible distributions of losses,
and it is referred to as an \emph{ambiguity set}.
To compare two ambiguity sets $\mathcal M_1,\mathcal M_2\subseteq \mathcal M(\R)$ in our context, two immediate observations are helpful. 
\begin{enumerate}
    \item Between two sets, a natural partial order is the set inclusion. If $\mathcal M_1\subseteq \mathcal M_2$, then $\mathcal M_2$ has more distribution models of losses, and thus it is more ambiguous.  
    \item The magnitude of losses in $\mathcal M_1$
and $\mathcal M_2$ also matters. For instance, if $\mathcal M_1$ contains only losses that take values in $[0,1]$
and $\mathcal M_2$ contains only losses that take values in $[2,3]$, then $\mathcal M_2$ is more dangerous, although the set inclusion $\mathcal M_1\subseteq \mathcal M_2$ does not hold.
\end{enumerate}
Motivated by these two considerations, we 
define a pre-order $\precsim$ by:
$\mathcal M_1\precsim \mathcal M_2$ 
if for any $F\in \mathcal M_1$, there exists $G\in\mathcal M_2$ such that $G\ge_{\rm st} F$. Note that the relation $\precsim $ is transitive and it is weaker than the set inclusion $\subseteq$.  
Thus, $\M_2$ is ``worse'' than $\M_1$ in the sense that every loss distribution in $\M_1$ is first-order stochastically smaller than some loss distribution in
$\M_2$.  


\begin{proposition}
\label{prop:1}
For two 
{nonempty, convex and weakly compact} $\mathcal M_1,\mathcal M_2\subseteq \mathcal M(\R)$, the following statements hold:
\begin{enumerate}[(i)]
    \item $\mathcal M_1\precsim \mathcal M_2$
if and only if 
$$
\sup_{\tilde x\lawis F\in \mathcal M_2} \E[\ell(\tilde x)] 
\ge 
\sup_{\tilde x\lawis  F\in \mathcal M_1} \E[\ell(\tilde x)] 
$$
for every increasing $\ell:\R\to\R$ such that the two suprema are well defined; 
    \item $\mathcal M_1\subseteq \mathcal M_2$
if and only if 
$$
\sup_{\tilde x\lawis F\in \mathcal M_2} \E[\ell(\tilde x)] 
\ge 
\sup_{\tilde x\lawis  F\in \mathcal M_1} \E[\ell(\tilde x)] 
$$
for every 
{Borel measurable} $\ell:\R\to\R$ such that the two suprema are well defined.
\end{enumerate}
\end{proposition}
Proposition \ref{prop:1} thus clarifies the natural interpretation of our notion of ambiguity comparison, which is very similar to that of set inclusion.  For two 
{nonempty, convex and weakly compact} ambiguity sets, the comparison $\mathcal M_1\precsim \mathcal M_2$ precisely means that any decision maker with monotonic utility over losses in the framework of maxmin expected utility of \cite{GS89} will prefer $\mathcal M_1$ over $\mathcal M_2$, noting that the increasing loss function $\ell$ can be interpreted as a disutility function. 

After introducing the degree of aggregate ambiguity, we will relate equity to ambiguity comparison. 

\begin{definition}[Aggregate ambiguity]\label{def:agg-ambiguity}
Let $\bm \theta, \bm \gamma\in\mathcal{A}(\bar\theta)$ for some total amount of resource $\bar \theta$ and $H$ be an aggregation function.
We say that allocation $\bm \theta$ has \emph{more aggregate ambiguity} than $\bm \gamma$  if
$
\D_H(\bm \gamma) \precsim \D_H(\bm \theta).
$
\end{definition}

\subsection{
A fundamental conflict}%

Given a resource level $\theta \in \Theta$, 
an individual $i$ incurs a loss $\tilde x_i$ following the distribution function (cdf) denoted by $F_\theta$. We make the following assumption on $F_\theta$, which is important for our analysis.

\begin{assumption}
\label{assm:convex}
    The mapping $\theta \mapsto F_\theta $ is convex on a convex subset $\Theta \subseteq \R$.
\end{assumption}

The convexity of $\theta \mapsto F_\theta$ in 
Assumption \ref{assm:convex}  has clear practical meaning: as resources increase, their protective benefit accelerates. 
If $\theta$ represents resource allocation, convexity of $\theta \mapsto F_\theta$ means that marginal benefit increases:
the gain in survival probability from the $k$-th unit of resources exceeds the gain from the $(k-1)$-th unit. 
This captures settings where resources exhibit increasing returns, for example, when investments unlock synergies, network effects, or critical thresholds beyond which protection becomes vastly more effective. For example, in pandemic control, vaccination exhibits strong increasing returns due to herd immunity. Initially, vaccines provide individual protection but limited population-level benefit. Once vaccination crosses a critical threshold (typically 60--70\%), transmission collapses rapidly, and marginal doses become vastly more effective. 






Below we give some examples satisfying Assumption \ref{assm:convex}.
\begin{example}
\label{ex:1}
Let  $G$ be a cdf on $\R$, $\Theta=[0,t)$ for some $t>0$,  and   $F_\theta(x)= G(x)^{t-\theta}$ for $x\in \R$, with the convention $0^0=0$.  
Then,  $\theta \mapsto F_\theta (x)$ is  convex at each $x\in \R$. 
\end{example} 

\begin{example}
\label{ex:2}
Let  $F_G$ and $F_B$ be two cdfs on $\R$ with $F_G\ge F_B$, $\Theta=[0,1]$,  and $w:[0,1]\to[0,1]$ is increasing and convex. 
Let $F_\theta(x)=w(\theta)  F_G(x)+ (1-w(\theta)) F_B(x)$ for $x\in \R$.
Then,  $\theta \mapsto F_\theta (x)$ is  convex at each $x\in \R$.
In particular, when $w$ is the identity, we have 
$F_\theta(x)= \theta  F_G(x)+ (1-\theta) F_B(x)$ and $\theta \mapsto F_\theta$ is affine.
\end{example} 

\begin{example}
\label{ex:3}
Let   $F_\theta=B_{1-\theta}$, where $B_p$ is the Bernoulli distribution with mean $p \in [0, 1]$. 
This corresponds to a special case of Example \ref{ex:2} with $F_G=\delta_0$, $F_B=\delta_1$, and that $w$ is the identity.
\end{example} 
Example~\ref{ex:1} illustrates the case where $\theta$ is a resource that reduces the loss
probability, so that a larger $\theta$ is more beneficial. When $m$ is a positive
integer and $G$ is the CDF of some loss $X$, the distribution $G(x)^m$ is exactly the CDF
of the maximum of $m$ i.i.d.~copies of $X$. In Example~\ref{ex:1} we have
$F_\theta(x)=G(x)^{t-\theta}$, which can be interpreted as follows (in the case that $t-\theta$ is an integer): suppose an individual
is exposed to $t-\theta$ independent ``rounds of being shot'', and let $G(x)$ denote the
probability that the loss in a single round does not exceed $x$. Then $G(x)^{t-\theta}$ is
the probability that the loss over all rounds is at most $x$. A larger $\theta$ corresponds
to fewer harmful rounds, and hence to a larger $F_\theta(x)$ for all $x$, that is, a smaller
loss in the usual stochastic order. This also aligns with the earthquake-resistant example: each dollar spent reduces collapse probability, but with increasing returns. Building codes that achieve high structural standards create positive spillovers: retrofitted buildings reduce cascading failures in earthquakes, making neighbors' buildings safer too. Initial spending on a few buildings helps only those structures; widespread adoption unlocks network benefits. The loss distribution improves convexly in aggregate spending on resilience. Formally, if $G$ is the baseline (no protection) loss distribution and $\theta$ is spending, a natural model is $F_\theta(x) = G(x)^{1-\alpha\theta}$ for small $\alpha > 0$. 

Example~\ref{ex:2} represents a case with one favorable distribution $F_G$ (good) and one less
preferred distribution $F_B$ (bad), and $w(\theta)$ is the mixing weight on the favorable distribution. As $\theta$ increases, $w(\theta)$ increases and the loss distribution moves from the less
preferred distribution $F_B$ towards the favorable distribution $F_G$. In particular, when
$w$ is the identity, we have $F_\theta(x) = \theta F_G(x) + (1-\theta)F_B(x)$, so that as
$\theta$ varies from $0$ to $1$, $F_\theta$ continuously switches from the bad distribution
to the good distribution.

The simple model in Example \ref{ex:3} will be a primary example in our paper, 
which is easy to interpret: If $\tilde x_i=1$ represents death and $\tilde x_i=0$ represents survival, then 
agent $i$ survives with probability $\theta$. Therefore, the resource $\theta$ represents the chance of survival, such as medical aids. This is the distribution model used by \cite{K80}, which will be further investigated in Section \ref{sec:keeney-model}.

Because it is better to have a small loss,   a bigger $F_\theta$ is more preferable to the agent and to the central planner. Therefore, 
when more resources are allocated to the agent, the agent's loss should naturally decrease, and its distribution function  should increase. Thus, it is natural  in applications to assume that $\theta \mapsto F_\theta $ is increasing (which is the case in Examples \ref{ex:1}--\ref{ex:3}). We do not need such an assumption  for our theoretical results. Moreover, as losses are typically nonnegative, it may be intuitively simpler to restrict the support of $F_\theta$ on $\R_+$; we do not need this assumption either.

\begin{theorem}
\label{th:1}
Assume that $\theta\mapsto F_\theta$ is convex and $H$ is increasing and symmetric. 
Let $\bm \theta,\bm \gamma\in \Theta^n$. 
If $\bm \theta\le_{\rm m} \bm \gamma$, then 
$
\D_H (\bm \gamma) \precsim \D _H (\bm \theta ).
$
\end{theorem}
\begin{proof}
Because  $\bm \theta\le_{\rm m} \bm \gamma$, there exists a doubly stochastic matrix $\Lambda\in \R^{n\times n}$ such that
$\bm \theta=\Lambda \bm \gamma$.
Let $$(G_1,\dots,G_n)^{\top} = \Lambda(F_{\gamma_1},\dots,F_{\gamma_n})^{\top}. $$
Note that since $\theta\mapsto F_{\theta}$ is convex, we have
$
F_{\theta_i}\le G_{i} 
$, that is, 
$
F_{\theta_i}\ge_{\rm st} G_{i} 
$, 
for all $i\in [n]$. 
Using Theorem 2.2 of \cite{CLLW22},
we get
\begin{equation}\label{eq:chen}
 \left\{ \text{cdf of } H(\tilde x_1,\dots,\tilde x_n): \tilde{x}_i \lawis  F_{\gamma_i}, \;i \in [n]\right\}
 \subseteq \left\{ \text{cdf of } H(\tilde x_1,\dots,\tilde x_n): \tilde{x}_i \lawis  G_{i}, \;i\in [n]\right\}.
\end{equation}
Therefore, 
\begin{align*}
    \mathcal D_H(\bm \gamma)&=  \left\{ \text{cdf of } H(\tilde x_1,\dots,\tilde x_n): \tilde{x}_i \lawis  F_{\gamma_i}, \;i \in [n]\right\}
\\&\precsim  \left\{ \text{cdf of } H(\tilde x_1,\dots,\tilde x_n): \tilde{x}_i \lawis  G_{i}, \;i\in [n]\right\}
\\& \precsim \left\{ \text{cdf of } H(\tilde x_1,\dots,\tilde x_n): \tilde{x}_i \lawis  F_{\theta_i}, \;i\in [n]\right\}
= \mathcal D_H(\bm \theta),
\end{align*}
where the last inequality follows from the increasing monotonicity of $H$ and the condition
$
F_{\theta_i}\ge_{\rm st} G_{i} 
$ 
for all $i\in [n]$. 
Therefore, by transitivity of $\precsim$, $  \mathcal D_H(\bm \gamma) \precsim \mathcal  D_H(\bm \theta)$. 
\end{proof}

In case $\theta\mapsto F_\theta$ is affine as in Example \ref{ex:2}, we have a slightly stronger conclusion on the comparison between  $\D_H (\bm \gamma) $ and  $ \D _H (\bm \theta ) $ without monotonicity of $H$. 

\begin{theorem}
\label{th:2}
{
Assume that $\theta\mapsto F_\theta$ is affine and $H$ is 
symmetric.} Let $\bm \theta,\bm \gamma\in \Theta^n$. 
If $\bm \theta\le_{\rm m} \bm \gamma$, then 
$
\D_H (\bm \gamma) \subseteq \D _H (\bm \theta).
$
\end{theorem}
\begin{proof}
     Let $U$ have the empirical law of $\bm\theta$ and $V$ that of $\bm\gamma$, giving each coordinate weight $1/n$. The relation $\bm\theta\le_m\bm\gamma$ says that $U$ is smaller than $V$ in convex order. The martingale coupling theorem of \cite{Str65} gives a joint law with $\E[V\mid U]=U$. Given $(U,V)$, independently draw $I$ uniformly from $\{i:\theta_i=U\}$ and $J$ uniformly from $\{j:\gamma_j=V\}$. Both indices are uniform on $[n]$, so the transition matrix $\Lambda_{ij}=\p(J=j\mid I=i)$ is doubly stochastic. Since the draw of $I$ uses only $U$,
\begin{equation*}
(\Lambda\bm\gamma)_i=\E[\gamma_J\mid I=i]=\E[V\mid U=\theta_i]=\theta_i.
\end{equation*}
We next use the permutation argument of \cite{CLLW22}. Take any random vector $\mathbf{Y}$ with marginals $F_{\gamma_i}$. Birkhoff's theorem (\citealp{MOA11}) represents $\Lambda$ as the mean of a random permutation matrix $\Pi$, which we choose independently of $\mathbf{Y}$. Then $\mathbf{X} = \Pi \mathbf{Y}$ has $i$th marginal cdf $\sum_j\Lambda_{ij}F_{\gamma_j}=F_{\theta_i}$ by affinity. Symmetry gives $H(\mathbf{X})=H(\mathbf{Y})$, so every law in $\D_H(\bm\gamma)$ also belongs to $\D_H(\bm\theta)$. 
\end{proof}



Combining Definition~\ref{def:resource-equity} with Theorems~\ref{th:1} and~\ref{th:2}, we immediately
obtain the following general message under the stated conditions: if $\bm \theta$ is more equitable than $\bm \gamma$,
that is, $\bm \theta \le_m \bm \gamma$, then
$$
\D_H(\bm \gamma)\precsim \D_H(\bm \theta),
$$
and in the affine case,  the stronger set inclusion $\D_H(\bm \gamma) \subseteq D_H(\bm \theta)$ holds. In other words, a move towards equity (replacing $\bm \gamma$ by a more equitable $\bm \theta$) enlarges the aggregation
set and increases aggregate ambiguity about the total loss.

A planner who tend to avoid {aggregate ambiguity}, in the sense of preferring allocations
with a smaller aggregation set (with respect to $\precsim$ or set inclusion), is therefore
in conflict with a strict preference for equity: more equitable allocations
necessarily come with more aggregate ambiguity about the total loss. This fundamental
trade-off will be made precise in the Bernoulli setting of Section \ref{sec:keeney-model}
and further analyzed in terms of $\alpha$-maxmin preferences in Section \ref{sec:preference}.

\section{The Keeney model}\label{sec:keeney-model}

\subsection{Setting}

We now connect to  the classic \cite{K80} model by   
using Example \ref{ex:3}.  In this model, $F_{\theta}$ becomes the Bernoulli cdf (hence, $\theta\mapsto F_\theta$ is affine) and the aggregation becomes the sum (i.e., $H$ is the sum function $\sum$).  

Formally, we consider a population of $n$ individuals potentially at risk, where $n \in \N$ is a positive integer. For $i\in [n]$, a resource $\theta_i \in \Theta = [0, 1]$ denotes the individual's survival probability (i.e., the individual $i$ faces a probability of death $1-\theta_i \in [0, 1]$). We model the risk of individual death $\tilde{x}_i$ as a Bernoulli random variable, i.e., $\tilde{x}_i \lawis B_{1-\theta_i}$, where $\tilde{x}_i$ takes the value 0 (survival) with probability $\theta_i$ and the value 1  (death) otherwise, and $B_q$ is a Bernoulli distribution with mean $q \in [0, 1]$. Assume that each individual is equally valued for the social planner. We are interested in the 
total number of deaths 
\begin{equation}\label{eq:Keeney-sum}
\tilde{d} = \sum_{i = 1}^n \tilde{x}_i.
\end{equation}
Again, the distribution of $\tilde{d}$ depends on not only the marginal distributions of all individual variables $\tilde{x}_1, \dots, \tilde{x}_n$, but also the dependence structure among them. 
If the dependence is completely specified, $\tilde{d}$ becomes a pure risk;   \cite{K80} assumed independence between the individual risks. But if the dependence is unknown, there is a great dependence uncertainty in the investigation of $\tilde{d}$, which is aggregate ambiguity in Definition \ref{def:agg-ambiguity}. In fact, this issue is more significant in practice, because it is difficult to characterize the dependence structure among individual losses, especially when involving high-dimensional circumstances.

The novelty of this section is to investigate the relationship between aggregate ambiguity and equity by obtaining stronger results than Theorems~\ref{th:1}--\ref{th:2} under this model. 
Again, we represent the possible distributions of total death by using the {aggregation set} of marginal distributions. The {profile} of the marginal Bernoulli distributions is denoted by a parameter vector $(\theta_1, \dots, \theta_n) \in [0, 1]^n$. We define the corresponding \textit{sum aggregation set} by
\begin{equation}\label{eq:aggregation}
\D_{\mathrm{sum}} (\theta_1, \dots, \theta_n) = \left\{ \text{cdf of } \sum_{i = 1}^{n} \tilde{x}_i:\; \tilde{x}_i \lawis  B_{1-\theta_i}, \;i = 1, \dots, n\right\}.
\end{equation} 
 {Thus $\D_{\mathrm{sum}}(\bm\gamma)\subseteq\D_{\mathrm{sum}}(\bm\theta)$ means that $\bm\theta$ can reproduce every distribution of total fatalities possible under $\bm\gamma$, even if the joint distribution of individual deaths differs.
{The Bernoulli aggregation sets are convex and weakly compact, since they are linear images of a compact convex set of joint probability mass vectors. Hence Proposition~\ref{prop:1} applies.}}

We will use increasing convex order to model the comparison of risk distributions. The notion of \textit{catastrophe aversion} defined in \cite{K80} is a special case where the random variables in comparison have only two outcomes.

\begin{definition}[Risk comparison]\label{def:catastrophic}
For two random variables $\tilde{d}$ and $\tilde{d}'$ representing losses with the same mean, $\tilde d$ is more \emph{risky} than $\tilde d'$ if $\tilde{d}$ is larger than $\tilde{d}'$ in \emph{increasing convex order}, that is, for any increasing convex function $f$, $\E[f(\tilde{d}')] \leq \E[f(\tilde{d})]$. We denote this case by $\tilde{d}' \ssd \tilde{d}$. In some places, we denote this case by $F \ssd G$, where $F$ is the distribution of $\tilde{d}'$ and $G$ is the distribution of $\tilde{d}$.
\end{definition}

If we use the terminology of \cite{AHT14} and   \cite{BRT18},
then the distribution of $\tilde{d}$ is  \emph{more catastrophic than} that of $\tilde{d}'$ if the order in Definition \ref{def:catastrophic} holds. 
Increasing convex order over losses can be converted to the classic second-order stochastic dominance over gains, with the same interpretation of preferring more sure wealth (less sure loss) and less distributional spread.

 Many early results on the Keeney model require the independence assumption; see \cite{K80,K80MS} and \cite{KW85}. This assumption may not hold in many real applications, and it significantly reduces the complexity of the setting. 
\cite{KW85} distinguish ex ante risk equity, concerning the fairness of the process that assigns risks to individuals, from ex post risk equity, concerning the distribution of eventual fatalities. In our model, equity is assessed through individuals' marginal survival probabilities and is therefore an ex ante criterion, regardless of the dependence structure among individual losses.
Further, \cite{F84} suggested that the same risk profile may correspond to multiple kinds of probability distributions due to the interdependence. \cite{BRT18} investigated the two-agent setting and analyzed the equity results in all kinds of correlation parameters.  Our treatment of dependence is  through ambiguity, different from above.
This requires models different from the expected utility model used in e.g., \cite{K80MS} and \cite{B82},
and an ambiguity preferences model will be analyzed in Section~\ref{sec:preference}.

%





\subsection{Equity and aggregate ambiguity in the Keeney model} \label{sec:aggregate}


Theorem \ref{thm:equiv}, our main result under the reduced Keeney model, characterizes the equivalence between equity and aggregate ambiguity:
Under Bernoulli marginals, equal treatment of risk profiles is precisely equivalent to an expansion of the uncertainty set for all feasible aggregate-loss distributions.
\begin{theorem}\label{thm:equiv}
Under the Keeney model \eqref{eq:Keeney-sum}-\eqref{eq:aggregation}, for $\bm \theta, \bm \gamma \in [0,1]^n$, the following are equivalent:
\begin{enumerate}[(i)]
\item $\bm \theta \le_m \bm \gamma$ (i.e., $\bm \theta$ is more equitable than $\bm \gamma$);
\item $\D_{\mathrm{sum}}(\bm \gamma)\subseteq \D_{\mathrm{sum}}(\bm \theta)$.  
\end{enumerate}
\end{theorem}

\begin{proof}
	The implication (i)$\Rightarrow$(ii) follows from Theorem \ref{th:2}. 
    Next we show  (ii)$\Rightarrow$(i). It suffices to show that $\bm \theta$ is dominated by $\bm \gamma$ in majorization order.
    Take comonotonic\footnote{The random variables $\tilde{x}_1,\ldots,\tilde{x}_n$  are called comonotonic, if there exists a random variable $\tilde{z}$ such that $\tilde{x}_i=h_i(\tilde{z})$ almost surely, $i\in [n]$, where $h_1,\ldots,h_n$ are non-decreasing functions. } random variables $\tilde{x}_i^c\lawis B_{1-\gamma_i},~i=1,\dots,n$ and let $\tilde d^c =\sum_{i=1}^n \tilde{x}_i^c$.
	For each $j\in \{1,\dots,n\}$, we can compute
	$$
	\E[(\tilde d^c -j)_+]  
	= \sum_{i=j+1}^{n} \p(\tilde d^c  \ge i)= \sum_{i=j+1}^{n}\p(\tilde{x}_{(i)}^c =1)
	=  \sum_{i=j+1}^{n} (1-\gamma_{[i]}),
	$$
	where $1-\gamma_{[i]}$ is the $i$-th largest order statistic of $1-\gamma_1,\dots,1-\gamma_n$, $\tilde{x}_{(i)}^c$  is the $i$-th largest order statistic of $\tilde x_1,\dots,\tilde x_n$, and the last equality follows from $\tilde{x}_{(i)}^c\lawis B_{1-\gamma_{[i]}}$ as $\tilde{x}_i ,~i=1,\dots,n$, are comonotonic. 
	Convexity of the function $x\mapsto (x-j)_+$ implies that comonotonicity maximizes its expectation  (see e.g., Theorem 3.5 of \citealp{R13}), that is, 
	$$\sum_{i=j+1}^{n} (1-\gamma_{[i]}) = \sup_{F\in \D_{\mathrm{sum}}(\bm \gamma)} \int_j^\infty (x-j) \d F(x). 
    $$
    Similarly,
    $$
    \sum_{i=j+1}^{n} (1-\theta_{[i]}) = \sup_{F\in \D_{\mathrm{sum}}(\bm \theta)} \int_j^\infty (x-j) \d F(x).
    $$
	Using 
    {$\D_{\mathrm{sum}}(\bm \gamma)\subseteq \D_{\mathrm{sum}}(\bm \theta)$}, this yields that the cdf of $\tilde d^c$ is also in {$\D_{\mathrm{sum}}(\bm \theta)$}
    and thus,
	\begin{equation}\label{eq:major1}
		\sum_{i=j+1}^{n} (1-\gamma_{[i]}) \le \sum_{i=j+1}^{n} (1-\theta_{[i]}) \mbox{~~~~for each $j\in \{1,\dots,n\}$.}
	\end{equation}
	Moreover, since the distributions in the set 
    {$\D_{\mathrm{sum}}(\bm \theta)$} 
    have the same mean and {$\D_{\mathrm{sum}}(\bm \gamma)\subseteq \D_{\mathrm{sum}}(\bm \theta)$}, 
    we  have
	\begin{equation}\label{eq:major2}  
		\sum_{i=1}^{n} (1-\gamma_{[i]})  =  \sum_{i=1}^{n} (1-\theta_{[i]}). 
	\end{equation}
Subtracting the sums of the $n-j$ smallest failure probabilities in \eqref{eq:major1} from the common total in \eqref{eq:major2} gives the reverse inequality for the sums of the $j$ largest. Their complements are the $j$ smallest survival probabilities, whose sum is therefore at least as large under $\bm\theta$ as under $\bm\gamma$. This gives $\bm\theta\le_m\bm\gamma$ by the conditions in Section~1.A.3 of \cite{MOA11}. 
\end{proof}

We give a simple example to illustrate the result in Theorem \ref{thm:equiv} that more equity implies more aggregate ambiguity.
\begin{example}\label{ex:simple}
Take $\bm \theta=(1/2,1/2)$ and $\bm \gamma=(0,1)$. We have $\bm \theta$ is more equitable than $\bm \gamma$. 
Denote by $\delta_{x}$ the point-mass at $x \in \R$. We have 
$$
\D_{\mathrm{sum}} (\theta_1,\theta_2)=\{s\delta_0+(1-2s)\delta_1+s\delta_2: s\in [0,1/2]\}~~~{\rm and}~~~ \D_{\mathrm{sum}} (\gamma_1,\gamma_2) = \left\{\delta_1\right\}.
$$
Obviously, $\D_{\mathrm{sum}} (\gamma_1,\gamma_2) \subseteq \D_{\mathrm{sum}} (\theta_1,\theta_2)$. 
\end{example}

The connection between equity, aggregate ambiguity, and  risk comparison is given in the next proposition. The direction (i)$\Rightarrow$(iii) is  essentially given in Theorem 1 of \cite{K80}, and we now show that the opposite direction does not hold, to highlight the difference between the concepts of aggregate ambiguity, equity, and riskiness.
\begin{proposition}\label{prop:2}
Under the Keeney model \eqref{eq:Keeney-sum}-\eqref{eq:aggregation}, the statements (i)-(ii) in Theorem \ref{thm:equiv} imply the following statement (iii), but (iii) does not generally imply (i)-(ii).
\begin{enumerate}[(iii)]
\item[(iii)] Given that individual risks are independent, the distribution of fatalities deduced by $\bm \theta$ is riskier than that deduced by $\bm \gamma$.
\end{enumerate}
\end{proposition}

\begin{proof}
	First, we show (i)$\Rightarrow$(iii). This statement is essentially given in Theorem 1 of \cite{K80}. We assume that $\bm \theta$ can be obtained from $\bm \gamma$ via a {Pigou--Dalton} transfer.
    Define $\mathbf{q} = (1-\theta_1, \dots, 1-\theta_n)$ and $\mathbf{p} = (1-\gamma_1, \dots, 1-\gamma_n)$. 
	Hence $\mathbf q$ can be obtained from $\mathbf p$ via a {Pigou--Dalton} transfer. As Definition 
    \ref{def:resource-equity}, without loss of generality, we denote $\pbf = (p_1, p_2, p_3, \dots, p_n)$ and $\qbf = ((p_1+\delta), (p_2-\delta), p_3, \dots, p_n)$, where $p_1 < p_2$ and $\delta \in (0, (p_2-p_1)/2]$. Denote by $\tilde{d} = \sum_{i = 1}^n \tilde{x}_i$ (resp., $\tilde{d}' = \sum_{i = 1}^{n} \tilde{y}_i$) the sum variable of fatalities deduced by $\mathbf{p}$ (resp., $\qbf$). We proceed to prove $\tilde{d} \ssd \tilde{d}'$. By direct computation, we obtain that
	$$
	\begin{aligned}
		& \p(\tilde{y}_1 + \tilde{y}_2 = 2) = (p_1 + \delta) (p_2 - \delta) > p_1 p_2 = \p(\tilde{x}_1 + \tilde{x}_2 = 2),\\
		& \p(\tilde{y}_1 + \tilde{y}_2 = 0) = (1 - p_1 - \delta) (1 - p_2 + \delta) > (1 - p_1) (1 - p_2) = \p(\tilde{x}_1 + \tilde{x}_2 = 0),\\
		& \p(\tilde{y}_1 + \tilde{y}_2 = 1) < \p(\tilde{x}_1 + \tilde{x}_2 = 1).
	\end{aligned}
	$$
	Hence, we have $(\tilde{x}_1 + \tilde{x}_2) \ssd (\tilde{y}_1 + \tilde{y}_2)$. Because of independence between marginals, we have
	$$
	\tilde{d} = \left( (\tilde{x}_1 + \tilde{x}_2) + \sum_{i = 3}^{n} \tilde{x}_i \right) \ssd \left( (\tilde{y}_1 + \tilde{y}_2) + \sum_{i = 3}^{n} \tilde{y}_i \right) = \tilde{d}'.
	$$
	Now for the general case that $\mathbf{q}$ is more equitable than $\mathbf{p}$, we know that $\mathbf q$ can be obtained from $\mathbf p$ via several {Pigou--Dalton} transfers and a permutation. Each Pigou--Dalton transfer will make the distribution of fatalities larger in increasing convex order than before and a permutation does not change the distribution of fatalities. Hence, we prove (iii).
	
	Finally, we propose a counter-example that (iii)$\Rightarrow$(i) does not hold. Let
	$$
	\begin{aligned}
		\pbf = \left(\frac{1}{2}, \frac{1}{2}, 0\right),\;\;
		\qbf = \left(\frac{3}{5}, \frac{1}{5}, \frac{1}{5}\right).
	\end{aligned}
	$$
	We know that $\mathbf{q}$ is NOT more equitable than $\mathbf{p}$ (also, $\mathbf{p}$ is NOT more equitable than $\mathbf{q}$). Denote by $\tilde{d} = \sum_{i = 1}^n \tilde{x}_i$ (resp., $\tilde{d}' = \sum_{i = 1}^{n} \tilde{y}_i$) the sum variable of fatalities deduced by $\mathbf{p}$ (resp., $\qbf$). By direct computation on the independent marginal variables, we obtain that
	$$
	\begin{alignedat}{2}
		& \p(\tilde{d} = 3) = 0, \;\; \p(\tilde{d}' = 3) = \frac{3}{125};
        &\qquad& \p(\tilde{d} = 2) = \frac{1}{4}, \;\; \p(\tilde{d}' = 2) = \frac{26}{125};\\
		& \p(\tilde{d} = 1) = \frac{1}{2}, \;\; \p(\tilde{d}' = 1) = \frac{64}{125};
        &\qquad& \p(\tilde{d} = 0) = \frac{1}{4}, \;\; \p(\tilde{d}' = 0) = \frac{32}{125}.
	\end{alignedat}
	$$
	It suffices to check that
	$$
	\begin{alignedat}{2}
		& \E[(\tilde{d} - 3)_+] = 0 = \E[(\tilde{d}' - 3)_+],
        &\qquad& \E[(\tilde{d} - 2)_+] = 0 < \frac{3}{125} = \E[(\tilde{d}' - 2)_+],\\
		& \E[(\tilde{d} - 1)_+] = \frac{1}{4} < \frac{32}{125} = \E[(\tilde{d}' - 1)_+],
        &\qquad& \E[(\tilde{d} - 0)_+] = 1 = \E[(\tilde{d}' - 0)_+].
	\end{alignedat}
	$$
	Hence, we have $\tilde{d} \ssd \tilde{d}'$, but $\mathbf{q}$ is NOT more equitable than $\mathbf{p}$, which means that (iii)$\Rightarrow$(i) does not hold.
\end{proof}

In \cite{K80}, the conflict between equity and catastrophe aversion  depends crucially on the independence assumption. In contrast, we show in Theorem \ref{thm:equiv} that the two properties, equity and aggregate ambiguity, have a fundamental equivalence which goes beyond the scope of dependence structure (i.e., not assuming independence). Further, Proposition \ref{prop:2} shows that greater aggregate risk  under independence is implied by the two properties. In the proof, we find a counterexample showing the inverse implication does not generally hold.  



In conjunction with Theorem \ref{thm:equiv}, we further have the following equivalence for general aggregation functions. 
\begin{theorem}\label{thm:equiv2}
Under the Keeney model \eqref{eq:Keeney-sum}-\eqref{eq:aggregation}, for $\bm \gamma, \bm \theta \in [0,1]^n$, the following are equivalent:
\begin{enumerate}[(i)]
    \item $\bm \theta \le_m \bm \gamma$ (i.e., $\bm \theta$ is more equitable than $\bm \gamma$);
	\item $\D_H(\bm \gamma)\subseteq \D_H(\bm \theta)$  for any symmetric aggregation function $H$. 
\end{enumerate}
\end{theorem}
\begin{proof}
	First, (i)$\Rightarrow$(ii) follows from Theorem \ref{th:2}. Next, (ii)$\Rightarrow$(i) holds. Indeed, we take $H$ as the sum function by the arbitrariness of $H$. The result holds if we use the proof logic similar to Theorem \ref{thm:equiv}(ii)$\Rightarrow$Theorem \ref{thm:equiv}(i).
\end{proof}

In \cite{K80}, the conflict between equity and catastrophe aversion  is not entirely due to  the independence assumption, but also lies in the sum aggregation. Indeed, if the social planner aggregates the risk in another form, the conflict may not exist. In contrast, we show in Theorem~\ref{thm:equiv2} that the fundamental conflict between equity and ($H$-specific) aggregate ambiguity avoidance not only goes beyond the scope of the dependence structure, but also holds for various kinds of aggregation function $H$. More discussions on this issue are given in Section \ref{sec:general}.

\section{Analysis under $\alpha$-maxmin model}
\label{sec:preference}

So far, we have been using ambiguity sets to make comparisons. Next we would like to adopt a quantitative model to describe the central planner's preferences for ambiguity. For this, we use
the $\alpha$-maxmin preferences (\citealp{GMM04}).
In contrast to other classic ambiguity models such as the smooth ambiguity model \citep{KMM05}  
and the variational model \citep{MMR06}, 
the $\alpha$-maxmin model directly takes the ambiguity set as input, and it does not require the specification of a prior or a penalty function on the ambiguity set. Hence it is natural to apply in our setting.

\subsection{The model}
Denote by $\mathcal M$ the collection of all distributions on $\R$. Fix some $\alpha \in [0, 1]$ and some 
utility function $u: \R \to \R$. 
The $\alpha$-maxmin preferences model is represented by maximizing the functional 
\begin{equation}\label{eq:alpha-maxmin}
\Phi_\alpha(\D) = \alpha \inf_{F \in \D} \int u \d F + (1 - \alpha) \sup_{F \in \D} \int u \d F, \quad \D \subseteq \mathcal{M}.
\end{equation}
The best-case $(\alpha = 0)$ and worst-case $(\alpha = 1)$ preferences are defined as
$$
\Phi_{0}(\D) = \sup_{F \in \D} \int u \d F\;\;\text{and} \;\; \Phi_{1}(\D) = \inf_{F \in \D} \int u \d F,\quad \D \subseteq \mathcal{M}.
$$
Intuitively, $\Phi_{0}$ represents ambiguity proneness and $\Phi_{1}$ represents ambiguity aversion. 

{Without loss of generality, assume that $u$ is not a constant function to avoid trivial cases.}  
For a symmetric aggregation function $H$, 
a decision maker with preference function $\Phi_{\alpha}$ prefers $\boldsymbol{\gamma} \in [0,1]^n$ to $\boldsymbol{\theta}\in [0,1]^n$ if
$$
\Phi_{\alpha}(\D_H(\boldsymbol{\gamma})) \ge \Phi_{\alpha}(\D_H(\boldsymbol{\theta})).
$$
We discuss two important criteria of the social planner. 

\begin{definition} \label{ass:RE}
For a given decision maker,  we say that 
\begin{enumerate}[(a)]
    \item  \emph{equity} holds if a more equitable vector of profiles is preferred to a less equitable vector; 
 
 \item 
 \emph{aggregate ambiguity avoidance (AAA)} holds for a fixed aggregation function $H$ if, for all feasible resource profiles $\boldsymbol{\gamma}, \boldsymbol{\theta} \in [0,1]^n$ satisfying $\sum_{i=1}^n \gamma_i = \sum_{i=1}^n \theta_i$, 
 $$
 \D_H(\boldsymbol{\gamma}) \subseteq \D_H(\boldsymbol{\theta}) \quad \Longrightarrow \quad \Phi_{\alpha}(\D_H(\boldsymbol{\gamma})) \geq \Phi_{\alpha}(\D_H(\boldsymbol{\theta})).
 $$
 Thus, an allocation with a smaller aggregation set under ordinary set inclusion is preferred.
\end{enumerate} 
\end{definition}
Note that AAA uses the ordinary set inclusion instead of the stochastic preorder $\precsim$ introduced in Section \ref{sec:main_results}.
This distinction applies throughout Sections \ref{sec:preference}-\ref{sec:general}.

\subsection{Main characterization results}
The criteria of equity and aggregate ambiguity avoidance are characterized by the attitudes on risk and ambiguity of the $\alpha$-maxmin preference relation,  shown in Theorem \ref{thm:ambi}.
Recall that all the comparisons in our setting are made for
all $n \in \N$ and all $\boldsymbol{\gamma}, \boldsymbol{\theta} \in [0, 1]^n$ satisfying
$\sum_{i = 1}^n \gamma_i = \sum_{i = 1}^n \theta_i$.
Further, under the Keeney model \eqref{eq:Keeney-sum}-\eqref{eq:aggregation}, the domain of $u$ should be $\mathbb{N} = \{0, 1, 2, \dots\}$ and hence convexity/concavity/affinity of $u$ should be understood in the discrete sense. In particular, we consider 
\begin{align*}
  &\text{discrete convexity:} 
  && u(k+2)-2u(k+1)+u(k)\ge 0,\qquad k=0,1,2,\dots.
\end{align*}

\begin{theorem}\label{thm:ambi}
Under the Keeney model \eqref{eq:Keeney-sum}-\eqref{eq:aggregation}, 
for an $\alpha$-maxmin decision maker \eqref{eq:alpha-maxmin}, 
\begin{enumerate}[(i)]
\item equity holds if and only if ambiguity proneness ($\alpha = 0$) or risk proneness ($u$ is discretely convex); 
\item 
AAA holds if and only if ambiguity aversion ($\alpha = 1$) or risk aversion ($u$ is discretely concave); 
\item both equity and AAA hold if and only if the preference is degenerate.
\end{enumerate}
\end{theorem}
\begin{proof}
	\begin{enumerate}[(i)]
		\item For the ``if" statement, fix two vectors of profiles $\boldsymbol{\gamma}, \boldsymbol{\theta} \in [0, 1]^n$ and $\boldsymbol{\theta}$ is more equitable than $\boldsymbol{\gamma}$. We aim to show $\Phi_{\alpha}(\D_{\mathrm{sum}}(\boldsymbol{\theta}))\ge \Phi_{\alpha}(\D_{\mathrm{sum}}(\boldsymbol{\gamma}))  $ when $\alpha=0$ or  $u$ is (discretely) convex. 
		By Theorem \ref{th:2}, we have   $\D_{\mathrm{sum}}(\boldsymbol{\gamma}) \subseteq \D_{\mathrm{sum}}(\boldsymbol{\theta})$. For any utility function $u$, we have
		\begin{eqnarray}\label{eq:proof_BC}
			\Phi_{0}(\D_{\mathrm{sum}}(\boldsymbol{\theta})) = \sup_{F \in \D_{\mathrm{sum}}(\boldsymbol{\theta})} \int u \d F \geq \sup_{F \in \D_{\mathrm{sum}}(\boldsymbol{\gamma})} \int u \d F = \Phi_{0}(\D_{\mathrm{sum}}(\boldsymbol{\gamma})).
		\end{eqnarray}
		This implies that equity holds for $\alpha = 0$. 
		We next consider the case that $u$ is convex. Let $\pbf = (1-\gamma_1, \dots, 1-\gamma_n)$ and $\qbf = (1-\theta_1, \dots, 1-\theta_n)$. Take $\tilde{x}_1^*\lawis B_{p_1}$, and iteratively let  $\tilde{x}_{i}^*\lawis B_{p_{i}}$ be countermonotonic with $\sum_{j=1}^{i-1} \tilde{x}_j^*$ for $i=2,\ldots,n$.  
		Denote by $F_0$ the cdf of $\sum_{i=1}^n\tilde{x}_i^*$. We have that $F_0\in \D_{\mathrm{sum}}(\boldsymbol{\gamma})(\subseteq \D_{\mathrm{sum}}(\boldsymbol{\theta}))$, and $F_0 \ssd F$ for any $F\in \D_{\mathrm{sum}}(\boldsymbol{\theta})$. As $u$ is convex, we have $\int u \d F_0 \le  \int u \d F$ for any $F\in \D_{\mathrm{sum}}(\boldsymbol{\theta})$, and thus,
		\begin{equation}\label{eq:convex}
			\Phi_{1}(\D_{\mathrm{sum}}(\boldsymbol{\theta})) = \inf_{F \in \D_{\mathrm{sum}}(\boldsymbol{\theta})} \int u \d F = \int u \d F_0 = \inf_{F \in \D_{\mathrm{sum}}(\boldsymbol{\gamma})} \int u \d F = \Phi_{1}(\D_{\mathrm{sum}}(\boldsymbol{\gamma})).
		\end{equation}
		This, together with \eqref{eq:proof_BC}, implies that for any $\alpha$,
		\begin{align*}
		  \Phi_{\alpha}(\D_{\mathrm{sum}}(\boldsymbol{\theta}))& = \alpha \Phi_{1}(\D_{\mathrm{sum}}(\boldsymbol{\theta})) + (1 - \alpha) \Phi_{0}(\D_{\mathrm{sum}}(\boldsymbol{\theta})) \\
          &\geq \alpha \Phi_{1}(\D_{\mathrm{sum}}(\boldsymbol{\gamma})) + (1 - \alpha) \Phi_{0}(\D_{\mathrm{sum}}(\boldsymbol{\gamma})) = \Phi_{\alpha}(\D_{\mathrm{sum}}(\boldsymbol{\gamma})).  
		\end{align*}
		
		For the ``only if" statement, it suffices to show that when $\alpha>0$, equity  implies that the preference is risk prone.
		We show this by contradiction. For $\alpha>0$, assume that $u$ is not discretely convex. There exists $m\in\N$ such that
		$$
		u(m+1) > \frac{1}{2}(u(m) + u(m+2)).
		$$
		Let $\boldsymbol{\gamma} = (1, 0, 0, \dots, 0)\in \R^{m+2}$ and $\boldsymbol{\theta} = \left(\frac{1}{2}, \frac{1}{2}, 0, \dots, 0\right)\in \R^{m+2}$. 
		As shown in Example 3.1 of \cite{MW15}, we have  $$\D_{\mathrm{sum}}(\boldsymbol{\gamma}) = \{\delta_{m+1}\}~~{\rm  and }~~\D_{\mathrm{sum}}(\boldsymbol{\theta})
=\{s\delta_{m}+(1-2s)\delta_{m+1}+s\delta_{m+2},~~s\in [0,1/2]\}.$$ 
		It follows that 
		\begin{equation}\label{eq:Phi_D_p}
			\begin{aligned}
				\Phi_{\alpha}(\D_{\mathrm{sum}}(\boldsymbol{\gamma})) &= \alpha \inf_{F \in \D_{\mathrm{sum}}(\boldsymbol{\gamma})} \int u \d F + (1 - \alpha) \sup_{F \in \D_{\mathrm{sum}}(\boldsymbol{\gamma})} \int u \d F
				= u({m+1}),
			\end{aligned}	
		\end{equation}
		and
		\begin{equation}\label{eq:Phi_D_q}
			\begin{aligned}
				\Phi_{\alpha}(\D_{\mathrm{sum}}(\boldsymbol{\theta}))
				& = \alpha \inf_{s\in [0,\frac{1}{2}]} \{s u(m) + (1-2s) u(m+1) + s u(m+2)\} \\
				&~~~+ (1 - \alpha) \sup_{s\in [0,\frac{1}{2}]}\{s u(m) + (1-2s) u(m+1) + s u(m+2)\}\\
				& = \alpha \frac{1}{2}(u(m) + u(m+2)) +(1 - \alpha)u(m+1),
			\end{aligned}	
		\end{equation}
		where the second equality follows from $\left(s u(m) + (1-2s) u(m+1) + s u(m+2)\right)$ is  decreasing in $s\in[0,1/2]$ as $u(m+1) > \frac{1}{2}(u(m) + u(m+2))$. Therefore, we have  $\Phi_{\alpha}(\D_{\mathrm{sum}}(\boldsymbol{\gamma})) > \Phi_{\alpha}(\D_{\mathrm{sum}}(\boldsymbol{\theta}))$ which yields a contradiction to equity. Thus, we have $u$ is convex when $\alpha>0$.
		
		\item The proof is similar to that of part (i).
		\item The result follows from (i) and (ii).\qedhere  
	\end{enumerate}
\end{proof}


We say that the $\alpha$-maxmin preference is \emph{degenerate} if for all $n \in \N$ and all $\boldsymbol{\gamma}, \boldsymbol{\theta} \in [0, 1]^n$ satisfying $\boldsymbol{\theta}$ is more equitable than $\boldsymbol{\gamma}$, 
we have
$\Phi_{\alpha}(\D_{H}(\boldsymbol{\gamma}))= \Phi_{\alpha}(\D_{H}(\boldsymbol{\theta}))$. 
That means, $\Phi_{\alpha}$ does not distinguish between any two profiles. 
This would in particular include the case that $u$ is affine; see Proposition \ref{prop:trivial}.
\begin{proposition}\label{prop:trivial}
Under the Keeney model \eqref{eq:Keeney-sum}-\eqref{eq:aggregation}, an $\alpha$-maxmin preference $\Phi_{\alpha}$ is degenerate if and only if it is one of the three cases:
\begin{enumerate}[(a)]
\item $u$ is affine on $\N$;
\item $u$ is discretely convex and $\alpha = 1$;
\item $u$ is discretely concave and $\alpha = 0$.
\end{enumerate}
\end{proposition}

\begin{proof}
	The ``if" side. If Case (a) holds, then the preference is degenerate. If Case (b) holds, we know from \eqref{eq:convex} that the preference is degenerate. Case (c) can be shown similarly.
	
	The ``only if" side. If $\alpha = 1$, we proceed to show that $u$ is (discretely) convex. Similarly to the proof of Theorem \ref{thm:ambi} (i), we prove by contradiction and assume that $u$ is not convex. The example in the proof of Theorem \ref{thm:ambi} (i) gives \eqref{eq:Phi_D_p}-\eqref{eq:Phi_D_q} and shows that $\Phi_{\alpha}$ is not degenerate, which is a contradiction.
	If $\alpha = 0$, we can show similarly that $u$ is (discretely) concave. If $\alpha \in (0, 1)$, we can show similarly that $u$ is affine (otherwise there is a contradiction due to \eqref{eq:Phi_D_p}-\eqref{eq:Phi_D_q}).
\end{proof}

Proposition \ref{prop:trivial} shows that the overlap of equity and aggregate ambiguity avoidance is genuinely ``degenerate". 
Indeed, the three cases in Proposition \ref{prop:trivial} are extreme. If $u$ is affine, then for two vectors $\boldsymbol{\gamma}$ and $\boldsymbol{\theta}$ with $\sum_{i = 1}^n \gamma_i = \sum_{i = 1}^n \theta_i$, we always have that $\boldsymbol{\gamma}$ and $\boldsymbol{\theta}$ are equally preferred, i.e., $\Phi_{\alpha}(\D_{\mathrm{sum}}(\boldsymbol{\gamma})) = \Phi_{\alpha}(\D_{\mathrm{sum}}(\boldsymbol{\theta}))$. This is a very special case. 
\subsection{Alternative ambiguity preference models}

The $\alpha$-maxmin framework is analytically tractable and yields clean characterizations. For alternative preference models, aggregate-set inclusion gives a direct comparison only when the evaluation rule is held fixed as the allocation changes. We describe the resulting comparisons and their limits.

\paragraph{Smooth ambiguity model.}

The smooth ambiguity model (\citealp{KMM05}) specifies preferences over ambiguity sets via a two-stage evaluation: the decision maker first computes expected utility under each distribution in the ambiguity set, and then applies a continuous and strictly increasing function $\phi$ to these expectations. Formally,
$$V(\mathcal{D}) = \int_{\mathcal{D}} \phi\left(\int u \, \d F\right) \d \mu(F),$$
where $\mu$ is a subjective probability over the ambiguity set and $\phi$ is the ambiguity attitude function. If $\phi$ is concave, then the decision maker is ambiguity-averse; if $\phi$ is convex, then the decision maker is ambiguity-prone.

In the Bernoulli setting with affine loss distributions (Theorem \ref{th:2}), equity implies $\mathcal{D}_H(\boldsymbol{\gamma}) \subseteq \mathcal{D}_H(\boldsymbol{\theta})$ when $\boldsymbol{\theta}$ is more equitable than $\boldsymbol{\gamma}$, meaning the range of expected utilities $\{\int u \, \d F : F \in \mathcal{D}_H(\boldsymbol{\theta})\}$ is larger. More generally, Theorem \ref{th:1} establishes the order $\mathcal{D}_H(\boldsymbol{\gamma}) \precsim \mathcal{D}_H(\boldsymbol{\theta})$. 
{Set inclusion alone does not determine this preference value, even when $\phi$ is concave. The second-order probability $\mu$ must also be specified for each allocation.}

The main distinction from $\alpha$-maxmin is that the interaction between risk attitude (convexity/\allowbreak concavity of $u$) and ambiguity attitude (convexity/\allowbreak concavity of $\phi$) is opaque. A complete characterization would require specifying all three components: $u$, $\phi$, and $\mu$.
It seems that there is no standard way of choosing different $\mu$   when the ambiguity set varies from $\mathcal{D}_H(\boldsymbol{\gamma})$ to $ \mathcal{D}_H(\boldsymbol{\theta})$.

\paragraph{Variational preferences.}

Variational preferences (\citealp{MMR06}) model ambiguity aversion through a penalty function:
$$V(\mathcal{D}) = \inf_{F \in \mathcal{D}} \left\{\int u \, \d F + c(F)\right\},$$
where $c(F)$ penalizes models that deviate from a baseline. 
With $\alpha = 1$ (worst-case), $\alpha$-maxmin coincides with  the special case $V(\mathcal{D}) = \inf_{F \in \mathcal{D}} \int u \, \d F$ with $c(F) = 0$, which is also the maxmin expected utility model of \cite{GS89}.

In the Bernoulli setting (Theorem \ref{th:2}), equity implies literal set inclusion $\mathcal{D}_H(\boldsymbol{\gamma}) \subseteq \mathcal{D}_H(\boldsymbol{\theta})$. 
{For the same fixed penalty $c$ on both sets, the infimum over the larger set is no greater, so $V(\mathcal{D}_H(\boldsymbol{\theta}))\le V(\mathcal{D}_H(\boldsymbol{\gamma}))$, whenever the values are well defined.}




\section{Numerical study}
\label{sec:numerical}

We illustrate our theoretical results through numerical examples that trace the equity-ambiguity tradeoff as resource allocations become progressively more equitable.

\subsection{Experimental setup}

We set $n = 3$ and construct a sequence of increasingly equitable resource profiles using an iterative matrix scheme. Define the initial death probability vector $\mathbf{p}^0 = (0.1, 0.5, 0.9)^\top$, representing a highly inequitable baseline. We then construct the transition matrix
\begin{equation}\label{eq:matrix}
\Lambda = 0.8 \times I_3 + 0.2 \times \left(\frac{1}{3}\right)_{3 \times 3},
\end{equation}
where $I_3$ is the $3 \times 3$ identity matrix and $\left(\frac{1}{3}\right)_{3 \times 3}$ denotes the $3 \times 3$ matrix of ones divided by 3. This structure smooths disparities while preserving the total: each iteration moves individuals closer to the mean death probability.

We define the sequence $\mathbf{p}^k := \Lambda^k \mathbf{p}^0$ for $k \in \mathbb{N}$. By construction, $\sum_{i=1}^3 p^k_i = \sum_{i=1}^3 p^0_i$ for all $k$ (resources are conserved), and $\mathbf{p}^k$ converges to $(1/2, 1/2, 1/2)^\top$ as $k \to \infty$ (perfect equity). The corresponding resource profile is $\boldsymbol{\gamma}^k := (1-p^k_1, 1-p^k_2, 1-p^k_3)^\top$, where $\gamma^k_i$ represents the survival probability for individual $i$ at iteration $k$. We track how preference values and inequity measures evolve as $k$ increases. 
To quantify the departure from equity, we employ two standard measures:
\begin{enumerate}
\item[(a)]  {Range penalty:} 
$$R(\boldsymbol{\gamma}) = \max_{i,j} |(1-\gamma_i) - (1-\gamma_j)|,$$
which captures the maximum difference in death probabilities across all pairs of individuals.

\item[(b)]  {Euclidean
penalty:} 
$$R(\boldsymbol{\gamma}) = \left\|\left(\mathbf{1}_n - \boldsymbol{\gamma}\right) - \frac{\sum_{i=1}^n (1-\gamma_i)}{n} \mathbf{1}_n\right\|_2,$$
which measures the $L_2$ distance of the death probability vector from its mean, capturing overall dispersion.
\end{enumerate}
As $k$ increases and profiles become more equitable (by majorization), both measures decrease toward zero.


For the aggregation set $\mathcal{D}(\boldsymbol{\gamma})$ under allocation $\boldsymbol{\gamma}$, we evaluate best-case and worst-case preferences:
$$\Phi_0(\mathcal{D}(\boldsymbol{\gamma})) = \sup_{F \in \mathcal{D}(\boldsymbol{\gamma})} \int u \, \d F \quad \text{and} \quad \Phi_1(\mathcal{D}(\boldsymbol{\gamma})) = \inf_{F \in \mathcal{D}(\boldsymbol{\gamma})} \int u \, \d F.$$
For concreteness, we use the utility function $u(x) = -x^2$ for $x \geq 0$. This function is concave (indicating risk aversion: marginal utility decreases as losses increase) and normalized at $u(0) = 0$ (zero deaths yield zero utility loss). The negative quadratic captures the intuition that society is increasingly averse to larger aggregate death counts. We compute $\Phi_0$ and $\Phi_1$ for each aggregation set $\mathcal{D}(\boldsymbol{\gamma}^k)$ as $k$ varies.

\subsection{Aggregation: Summation}

We first consider the classical sum aggregation, where total losses accumulate linearly. The aggregation set is
$$\mathcal{D}_{\mathrm{sum}}(\boldsymbol{\gamma}) = \left\{\text{cdf of } \sum_{i=1}^n \tilde{x}_i : \tilde{x}_i \sim B_{1-\gamma_i}, \; i = 1, \ldots, n\right\}.$$

Using Theorem \ref{thm:equiv2}, we can compute the $\alpha$-maxmin preference explicitly. The worst-case scenario corresponds to comonotonic dependence (
{the failure events are nested}), while the best case corresponds to 
{a coupling whose total number of failures takes only the two integers adjacent to its mean}. For $u(x) = -x^2$ and $(1 - \gamma_1) \leq (1 - \gamma_2) \leq (1 - \gamma_3)$, these extremes yield:
\begin{equation}\label{eq:alpha-maxmin-sum}
\begin{aligned}
\Phi_\alpha(\mathcal{D}_{\mathrm{sum}}(\boldsymbol{\gamma})) &= \alpha \underbrace{\left[u(3)(1-\gamma_1) + u(2)((1-\gamma_2) - (1-\gamma_1)) + u(1)((1-\gamma_3) - (1-\gamma_2)) + u(0)\gamma_3\right]}_{\text{worst-case (comonotonic)}}\\
&\quad + (1-\alpha) \underbrace{\left[u\left(\lceil \bar{d} \rceil\right) \cdot \{\bar{d}\} + u\left(\lfloor \bar{d} \rfloor\right) \cdot (1-\{\bar{d}\})\right]}_{\text{best-case (countermonotonic)}},
\end{aligned}
\end{equation}
where $\bar{d} = \sum_{i=1}^n (1-\gamma_i)$ is the expected number of failures and $\{\bar{d}\}$ denotes its fractional part. The worst-case utility decreases in equity (
{equalizing probabilities while keeping $\sum_i(1-\gamma_i)$ fixed}), while the best-case utility is constant across equitable profiles.

\begin{figure}[htbp]
\centering
\subfigure[Utility vs. Penalty Tradeoff]{\label{fig:1}
\begin{tikzpicture}
\begin{axis}[
width=0.47\textwidth,height=5.0cm,
xmin=0,xmax=30,ymin=-5,ymax=1,
xlabel={$k$},ylabel={{Penalty / utility}},
tick label style={font=\scriptsize},
label style={font=\small},
legend style={font=\scriptsize,nodes={inner ysep=0pt},at={(0.98,0.78)},anchor=north east},
grid=major,grid style={black!15},samples=16,domain=0:30]
\addplot[black,thick,mark=asterisk,mark size=1.5pt] {0.8*pow(0.8,x)};
\addlegendentry{Range penalty}
\addplot[red,thick,mark=triangle*,mark size=1.5pt] {0.4*sqrt(2)*pow(0.8,x)};
\addlegendentry{Euclidean penalty}
\addplot[blue,thick,dashed] {-4.5+1.6*pow(0.8,x)};
\addlegendentry{Worst-case utility}
\addplot[green!50!black,thick,dashed] {-2.5};
\addlegendentry{Best-case utility}
\end{axis}
\end{tikzpicture}}
\subfigure[Worst- vs Best-cases Tradeoff]{\label{fig:2}
\begin{tikzpicture}
\begin{axis}[
width=0.47\textwidth,height=5.0cm,
xmin=0,xmax=1,ymin=-4.7,ymax=-2.3,
xlabel={$\alpha$},ylabel={$\alpha$-maxmin utility},
tick label style={font=\scriptsize},
label style={font=\small},
grid=major,grid style={black!15},samples=2,domain=0:1]
\addplot[blue,thick] {-4.5*x-2.5*(1-x)};
\end{axis}
\end{tikzpicture}}
\caption{Numerical illustration of the equity-ambiguity tradeoff under sum aggregation.  {Left panel:} Inequity penalties (range and Euclidean) versus 
{best- and worst-case utilities} as profiles evolve toward equity ($k = 0, 2, 4, \ldots, 30$). Penalties decrease toward zero; worst-case utility decreases, confirming the conflict. 
{Best-case utility remains $-2.5$.} Inequity measures: Range penalty $= \max_{i,j} |p_i - p_j|$; Euclidean penalty $= \|(p_1, p_2, p_3) - \bar{p} \mathbf{1}_3\|_2$ where $\bar{p} = (p_1+p_2+p_3)/3$.  {Right panel:} 
{The $\alpha$-maxmin preference as a function of} $\alpha \in [0,1]$ for the limiting equitable profile $\mathbf{p} = (1/2, 1/2, 1/2)^\top$. The linear tradeoff $\Phi_\alpha = -4.5\alpha + (-2.5)(1-\alpha)$ 
{interpolates between the best-case value $-2.5$ at $\alpha=0$ and the worst-case value $-4.5$ at $\alpha=1$, holding the allocation fixed}.}
\label{fig:tradeoff}
\end{figure}
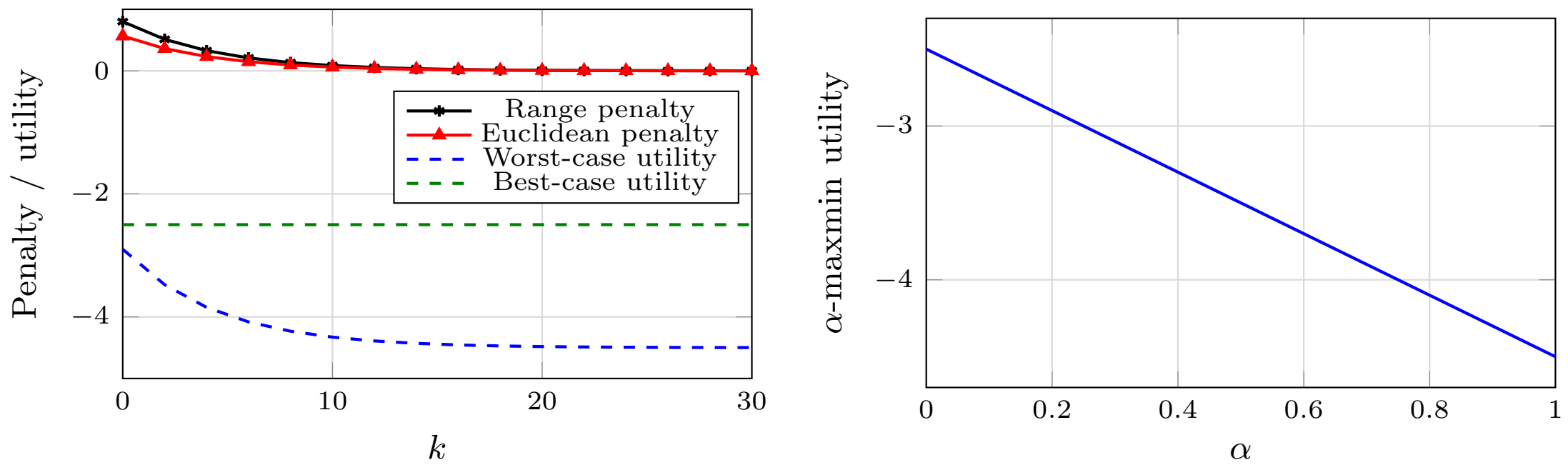

Figure \ref{fig:tradeoff} displays the results. We observe:

\begin{enumerate}
\item[(a)]   \textbf{Equity increases:}  As profiles become more equitable (increasing $k$), both the range and Euclidean penalties on inequity decrease to zero. This is expected: equity is measured by these penalties.

\item[(b)] \textbf{Equity-AAA conflict (worst-case):} The worst-case utility $\Phi_1(\mathcal{D}_{\mathrm{sum}}(\boldsymbol{\gamma}^k))$ strictly decreases as $k$ increases. This directly confirms Theorem \ref{thm:ambi}(ii): an ambiguity-averse planner ($\alpha=1$) faces worse outcomes under more equitable allocations, because equity expands the aggregation set and the comonotonic scenario (worst-case) becomes more severe. This is the core equity-AAA conflict.

\item[(c)] \textbf{Best-case:} The best-case utility $\Phi_0(\mathcal{D}_{\mathrm{sum}}(\boldsymbol{\gamma}^k))$ is invariant in $k$. The countermonotonic (best-case) distribution attains the same supremum regardless of how equitable the profile is, because perfect diversification always yields the same total expectation.

\item[(d)] \textbf{Intermediate cases ($0 < \alpha < 1$):} For intermediate values of $\alpha$, the preference value decreases with $k$, weighted between worst and best cases. A decision maker 
{placing equal weight on the best and worst cases ($\alpha = 1/2$)} faces a clear tradeoff between equity and ambiguity.
\end{enumerate}

These findings validate Theorem \ref{thm:ambi} and illustrate the quantitative magnitude of the equity-ambiguity conflict in the sum aggregation.

\subsection{Balancing equity and aggregate ambiguity avoidance: An optimization perspective}

To better understand the practical tradeoff between equity and aggregate ambiguity avoidance (AAA), we formulate an optimization problem that balances both objectives:
\begin{equation}\label{eq:obje}
\max_{\boldsymbol{\gamma} \in [0,1]^n: \sum_{i=1}^n (1-\gamma_i) = s} \left\{\inf_{F \in \mathcal{D}_{\mathrm{sum}}(\boldsymbol{\gamma})} \int u \d F - \lambda R(\boldsymbol{\gamma})\right\},
\end{equation}
where $\lambda > 0$ is a Lagrange multiplier that trades off worst-case utility against inequity. Larger $\lambda$ penalizes inequity more heavily, pushing solutions toward more equitable allocations; smaller $\lambda$ prioritizes worst-case protection. A related constrained formulation is
$$\max_{\boldsymbol{\gamma} \in [0,1]^n: \sum_{i=1}^n (1-\gamma_i) = s} \left\{\inf_{F \in \mathcal{D}_{\mathrm{sum}}(\boldsymbol{\gamma})} \int u  \d F : R(\boldsymbol{\gamma}) \leq c\right\}$$
for upper-bound constraint $c > 0$. Problem \eqref{eq:obje} allows us to trace the Pareto frontier between AAA and equity.

\begin{proposition}\label{prop:multiplier}
For $n = 3$ with $\gamma_1 \geq \gamma_2 \geq \gamma_3$ and concave utility $u(x) = -x^2$:

\begin{enumerate}
\item[(i)] The worst-case utility can be expressed as
$$\inf_{F \in \mathcal{D}_{\mathrm{sum}}(\boldsymbol{\gamma})} \int u \d F = u(3)(1-\gamma_1) + u(2)(\gamma_1 - \gamma_2) + u(1)(\gamma_2 - \gamma_3) + u(0)\gamma_3.$$
Problem \eqref{eq:obje} thus reduces to
\begin{equation}\label{prob:multiplier}
\max_{\boldsymbol{\gamma} \in [0,1]^3: \sum_{i=1}^3 (1-\gamma_i) = s} \left\{u(3)(1-\gamma_1) + u(2)(\gamma_1-\gamma_2) + u(1)(\gamma_2-\gamma_3) + u(0)\gamma_3 - \lambda R(\boldsymbol{\gamma})\right\}.
\end{equation}

\item[(ii)] Using the 
{squared} Euclidean penalty $R(\boldsymbol{\gamma}) = \left\|(\mathbf{1}_3 - \boldsymbol{\gamma}) - \frac{\sum_{i=1}^3(1-\gamma_i)}{3}\mathbf{1}_3\right\|_2^2$, the interior solution to Problem \eqref{prob:multiplier} has explicit form:
\begin{equation}\label{eq:solution}
\boldsymbol{\gamma}^* = \left(1 - \frac{s}{3} - \frac{2A - B - C}{6\lambda}, \; 1 - \frac{s}{3} - \frac{-A + 2B - C}{6\lambda}, \; 1 - \frac{s}{3} - \frac{-A - B + 2C}{6\lambda}\right),
\end{equation}
where $A = u(3) - u(2)$, $B = u(2) - u(1)$, $C = u(1) - u(0)$, given the necessary interior conditions hold: $\boldsymbol{\gamma}^* \in [0,1]^3$ and it is consistent with the assumed ordering $\gamma_1 \geq \gamma_2 \geq \gamma_3$.
\end{enumerate}
\end{proposition}

\begin{proof}
The reduction in part (i) follows from equation \eqref{eq:alpha-maxmin-sum} with $\alpha = 1$ (worst-case focus). Part (ii) is a direct application of first-order optimality conditions to the Lagrangian of Problem \eqref{prob:multiplier}. Details are standard and omitted.
\end{proof}

This explicit solution in Proposition \ref{prop:multiplier} characterizes how optimal allocations vary with the tradeoff parameter $\lambda$: as $\lambda$ increases (equity becomes more valued), the solution $\boldsymbol{\gamma}^*$ moves toward 
{$(1-s/3)\mathbf{1}_3$ (perfect equality)}.
This optimization framework operationalizes the theoretical conflict between equity and AAA, allowing practitioners to compute explicit Pareto-optimal allocations given their preference for each objective.

\subsection{Aggregation: Product}

We next consider a product aggregation, where the aggregate loss is the product of individual losses:
$$\mathcal{D}_{\text{prod}}(\boldsymbol{\gamma}) = \left\{\text{cdf of } \prod_{i=1}^n \tilde{x}_i : \tilde{x}_i \sim B_{1-\gamma_i}, \; i = 1, \ldots, n\right\}.$$

Since each $\tilde{x}_i \in \{0,1\}$ represents failure (1) or survival (0), the product $\prod_{i=1}^n \tilde{x}_i = 1$ only if \emph{all} individuals fail simultaneously. This aggregation models scenarios requiring universal failure, such as a redundant safety system where catastrophic loss occurs only if all backup mechanisms fail together. In contrast, the aggregation $H_a(\mathbf{x}) = 1 - \prod_{i=1}^n (1-x_i)$, which we examine in Section \ref{sec:general}, models the opposite extreme: aggregate loss occurs if \emph{any} individual fails.

For 
$$
\prod_{i=1}^n \tilde{x}_i =1
  \quad\Longleftrightarrow\quad
  \tilde x_1=\cdots=\tilde x_n=1,
$$  
the only relevant dependence quantity is
$s=P(\tilde x_1=\cdots=\tilde x_n=1)$, 
whose Fr\'{e}chet bounds are
$$
  s_{\min}=\max\left\{\sum_{i=1}^n (1-\gamma_i)-(n-1),0\right\} = \max\left\{\sum_{i=1}^3 (1-\gamma_i)-2,0\right\},
  \qquad
  s_{\max}=\min_i (1-\gamma_i) = 1-\gamma_1,
$$
because $(1 - \gamma_1) \leq (1 - \gamma_2) \leq (1 - \gamma_3)$. Using $u(x) = -x^2$, we compute:
\begin{equation*}
\Phi_\alpha(\mathcal{D}_{\text{prod}}(\boldsymbol{\gamma})) = \alpha \underbrace{[s_{\max}u(1)+(1-s_{\max})u(0)]}_{\text{worst-case}} + (1-\alpha) \underbrace{[s_{\min}u(1)+(1-s_{\min})u(0)]}_{\text{best-case}},
\end{equation*}
where the worst case is dominated by the minimum failure probability (the ``bottleneck" individual).

The qualitative behavior mirrors the sum case: worst-case utility decreases with equity (more equitable allocation makes simultaneous failure of all individuals more likely), while best-case utility remains 
{constant}. Since the product form yields similar qualitative insights without additional novelty, we omit detailed numerical results. The conflict between equity and ambiguity avoidance persists under this aggregation, though the specific mechanism differs: here, equity increases the 
{largest feasible} probability of total system failure, whereas under the sum aggregation, equity increases 
{the severity of the worst-case aggregate distribution while leaving the expected number of failures unchanged}.

\subsection{Aggregation: Elementary symmetric polynomials}

To demonstrate generality across diverse aggregation functions, we consider the second-order elementary symmetric polynomial:
$$\tilde{y} = \tilde{x}_1 \tilde{x}_2 + \tilde{x}_2 \tilde{x}_3 + \tilde{x}_1 \tilde{x}_3,$$
which captures scenarios where outcomes depend on pairwise interactions (e.g., cascading failures affecting pairs of adjacent systems).

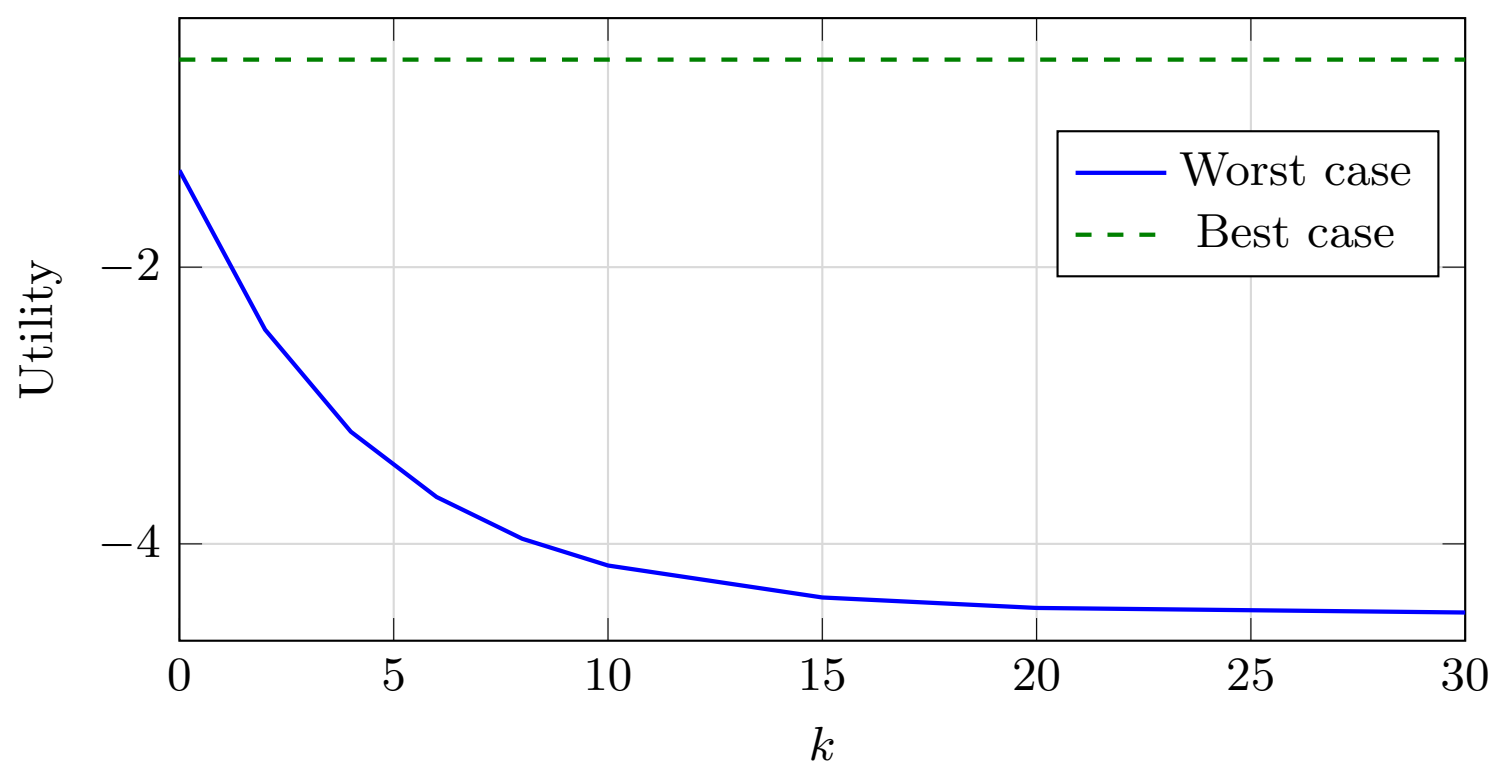
\begin{figure}[htbp]
\centering
\begin{tikzpicture}
\begin{axis}[
width=0.64\textwidth,height=5.7cm,
xmin=0,xmax=30,ymin=-4.7,ymax=-0.2,
xlabel={$k$},ylabel={Utility},
axis line style={black},tick style={black},
tick label style={font=\small,text=black},
label style={font=\small,text=black},
legend style={font=\small,text=black,draw=black,at={(0.98,0.82)},anchor=north east},
grid=major,grid style={black!15},samples at={0,2,4,6,8,10,15,20,30}]
\addplot[blue,thick] {-4.5+3.2*pow(0.8,x)};
\addlegendentry{Worst case}
\addplot[green!50!black,thick,dashed] {-0.5};
\addlegendentry{Best case}
\end{axis}
\end{tikzpicture}
\caption{Worst-case and best-case utilities under elementary symmetric aggregation as profiles evolve toward equity. Settings: $\mathbf{p}^k = \Lambda^k \mathbf{p}^0$, $k = 0, 2, 4, 6, 8, 10, 15, 20, 30$ (see \eqref{eq:matrix}). The downward trend in worst-case utility and the 
{constancy of} best-case utility illustrate the fundamental equity-AAA conflict across general aggregation functions.}
\label{fig:elem}
\end{figure}

The aggregation set is
$$\mathcal{D}_{\text{elem}}(\boldsymbol{\gamma}) = \left\{\text{cdf of } \tilde{x}_1 \tilde{x}_2 + \tilde{x}_2 \tilde{x}_3 + \tilde{x}_1 \tilde{x}_3 : \tilde{x}_i \sim B_{1-\gamma_i}, \; i = 1, 2, 3\right\}.$$
To compute the best and worst cases, we set $p_i = 1 - \gamma_i$, assume $p_1 \leq p_2 \leq p_3$ and define joint failure probabilities:
$$\pi_{ij} = {P}(\tilde{x}_i = 1, \tilde{x}_j = 1), \quad \pi_{123} = {P}(\tilde{x}_1 = 1, \tilde{x}_2 = 1, \tilde{x}_3 = 1).$$
The extremal cases reduce to solving a linear program:
\begin{equation}\label{eq:linprog-elem}
\begin{aligned}
\text{maximize/minimize} \quad &u(3)\pi_{123} + u(1)(\pi_{12} + \pi_{23} + \pi_{13} - 3\pi_{123}) \\
\text{subject to} 
\quad &\pi_{ij} \in [\max\{0, p_i + p_j - 1\}, \min\{p_i, p_j\}] \quad \text{for } ij \in \{12, 23, 13\},\\
&\pi_{123} \in [0, \min\{\pi_{12}, \pi_{23}, \pi_{13}\}],\\
&p_i - \pi_{123} - (\pi_{ij} + \pi_{ik} - 2\pi_{123}) \geq 0 \quad \text{for all distinct } i,j, k,\\
&
1-p_1-p_2-p_3+\pi_{12}+\pi_{13}+\pi_{23}-\pi_{123}\ge 0.
\end{aligned}
\end{equation}
Solving this program numerically for the sequence of profiles $\boldsymbol{\gamma}^k$ (as $k$ increases toward equity) yields Figure \ref{fig:elem}. The results reaffirm the central conflict: as profiles become more equitable ($k$ increases), the worst-case utility strictly decreases while the best-case utility 
{remains constant}. The aggregation set expands, creating a larger spread between optimistic and pessimistic scenarios. This demonstrates that the equity-AAA conflict is not an artifact of the sum aggregation but persists across diverse aggregation functions, 
{illustrating} Theorem~\rev{\ref{thm:equiv2}}.

\section{Further discussion on aggregation functions}\label{sec:general}

We extend the scope of the above findings to more symmetric aggregation functions in the fatality context (hence, $F_\theta$ becomes the Bernoulli cdf) under the $\alpha$-maxmin model \eqref{eq:alpha-maxmin}. Note that the Keeney model requires that the aggregation is a sum, as Eq. \eqref{eq:Keeney-sum} indicates. As a theoretical discussion, we also include the case where $u$ may not be decreasing.  

\subsection{
Any symmetric aggregation function}

If we take into consideration any (i.e., not necessarily increasing) symmetric aggregation function, we have the following characterization result.

\begin{theorem}\label{thm:all_H}
Assume $F_\theta=B_{1-\theta}$. For an $\alpha$-maxmin decision maker, 
\begin{enumerate}[(i)]
	\item equity holds for any symmetric aggregation function $H$ if and only if the preference is ambiguity prone ($\alpha = 0$); 
	\item 
    AAA holds for any symmetric aggregation function $H$ if and only if the preference is ambiguity averse ($\alpha = 1$); 
	\item 
		{No single preference satisfies both equity and AAA 
        for every symmetric aggregation function $H$}.
	\end{enumerate}
\end{theorem}

Theorem \ref{thm:all_H} shows that the essential feature of whether equity or AAA 
holds is the ambiguity attitude (i.e., $\alpha = 0$ or 1). The form of aggregation and other properties of a utility, including the risk attitude (convexity/concavity of $u$ and the monotonicity of $u$), may also have an impact on the decision criteria; e.g., in Theorem \ref{thm:ambi}. 
We will repeatedly see this point in the following characterization under other aggregation functions. 


\subsection{
Some examples}

First, we propose the following specific aggregation function
\begin{equation}\label{eq:multiply}
H_{\text{a}}(x_1, \dots, x_n) = 1 - \prod_{i = 1}^{n} (1-x_i),\qquad x_1,\dots,x_n\in [0,1].
\end{equation}
It means that any individual failure triggers system loss: if some individual is dead ($\tilde{x}_i = 1$), then the society will die ($H = 1$).  For example, this aggregation function is suitable to describe the decision towards an infectious virus or the situation where every life is of the utmost importance. 
The social planner is concerned about and trying to avoid any possible loss of the society.

Now we characterize the $\alpha$-maxmin preference for this aggregation function.

\begin{theorem}\label{thm:multiply}
Assume $F_\theta=B_{1-\theta}$.
For an $\alpha$-maxmin decision maker with the aggregation function \eqref{eq:multiply}, 
\begin{enumerate}[(i)]
	\item equity holds if and only if the preference is ambiguity prone ($\alpha = 0$) or satisfies $u(1) \leq u(0)$;
	\item AAA 
    holds if and only if the preference is ambiguity averse ($\alpha = 1$) or satisfies $u(1) \geq u(0)$;
	\item both equity and AAA 
    hold if and only if the preference satisfies: (a) $u(1) = u(0)$; 
		{or (b) ambiguity aversion ($\alpha = 1$) and $u(1) < u(0)$; or (c) ambiguity proneness ($\alpha = 0$) and $u(1) > u(0)$}.
	\end{enumerate}
\end{theorem}

As death always decreases utility ($u(1) < u(0)$), Theorem \ref{thm:multiply} shows that equity always holds under the aggregation function \eqref{eq:multiply}. The insight is intuitive: all individuals are highly dependent on and important to each other so that none of them should have a relatively high death rate. Thus, the survival rate for each individual should be as equalized as possible.


Next, we propose the following specific aggregation function
\begin{equation}\label{eq:multiply2}
	H_{\text{b}}(x_1, \dots, x_n) = \prod_{i = 1}^{n} x_i, \qquad x_1,\dots,x_n\in [0,1]
\end{equation}
This aggregation function is more suitable to describe the protection for some extremely endangered species, etc. If some individual is alive ($\tilde{x}_i = 0$), then the society will survive ($H_b = 0$). The planner is aiming to protect at least one individual in society.

\begin{theorem}\label{thm:multiply2}
Assume $F_\theta=B_{1-\theta}$.  For an $\alpha$-maxmin decision maker with the aggregation function \eqref{eq:multiply2}, 
	\begin{enumerate}[(i)]
		\item AAA 
        holds if and only if the preference is ambiguity averse ($\alpha = 1$) or satisfies $u(1) \leq u(0)$;
		\item equity holds if and only if the preference is ambiguity prone ($\alpha = 0$) or satisfies $u(1) \geq u(0)$;
		\item both equity and AAA 
        hold if and only if the preference satisfies: (a) $u(1) = u(0)$; or (b) ambiguity aversion ($\alpha = 1$) and $u(1) > u(0)$; or (c) ambiguity proneness ($\alpha = 0$) and $u(1) < u(0)$.
	\end{enumerate}
\end{theorem}

As death always decreases utility ($u(1) < u(0)$), Theorem \ref{thm:multiply2} shows that aggregate ambiguity avoidance always holds under the aggregation function \eqref{eq:multiply2}. 
Actually, in order to protect at least one individual, the social planner should provide all the effort to guarantee some individual's life and improve the corresponding survival probability as much as possible. 


\subsection{
General properties of aggregation functions: Sub/super-modularity}

Two aggregation functions \eqref{eq:multiply} and \eqref{eq:multiply2} actually correspond to mathematical properties of sub/super-modularity. In the following, we generalize the results in the previous section to the case of a general componentwise nondecreasing and submodular $H$. A function $H:\R^n\to \R$ is said to be submodular if
$$
H(\mathbf{x}) + H(\mathbf{y}) \ge  H(\mathbf{x} \vee  \mathbf{y}) + H(\mathbf{x} \wedge  \mathbf{y})~~~{\rm for~any~}\mathbf{x},  \mathbf{y}\in\R^n,
$$
where $\mathbf{x} \vee  \mathbf{y} = (\max\{x_1, y_1\}, \dots, \max\{x_n, y_n\})$ and $\mathbf{x} \wedge \mathbf{y} = (\min\{x_1, y_1\}, \dots, \min\{x_n, y_n\})$.  Denote by 
$$
\mathcal H_{\rm sub} := \{H: H\textrm{ is symmetric,  componentwise nondecreasing and submodular}\}.
$$ 
On the other hand, a function $H:\R^n\to \R$ is said to be supermodular if
$$
H(\mathbf{x}) + H(\mathbf{y}) \le  H(\mathbf{x} \vee  \mathbf{y}) + H(\mathbf{x} \wedge  \mathbf{y})~~~{\rm for~any~}\mathbf{x},  \mathbf{y}\in\R^n.
$$ 
For example, the aggregation function \eqref{eq:multiply} is submodular, while \eqref{eq:multiply2} is supermodular. The sub/super-modularity has a close connection to the equity axioms for public risk (\cite{F84}). 
The submodularity of $H$ implies the so-called \emph{common-fate equity} of \cite{F84}, while the supermodularity of $H$ implies the \emph{catastrophe-avoidance preference} of \cite{F84}. 

Now we have a characterization for equity and aggregate ambiguity avoidance under the $\alpha$-maxmin preference. Based on different sets of axioms, one can specify different classes of aggregation functions.
\begin{theorem}\label{thm:sub}
	Assume $F_\theta=B_{1-\theta}$. For an $\alpha$-maxmin decision maker,
	\begin{enumerate}[(i)]
		\item  equity holds for any $H\in \mathcal H_{\rm sub}$ if and only if the preference is ambiguity prone ($\alpha=0$) or $u$ is nonincreasing and convex;
		\item AAA 
        holds for any $H\in \mathcal H_{\rm sub}$ if and only if the preference is ambiguity averse ($\alpha=1$) or $u$ is nondecreasing and concave.
		\item both equity and AAA 
        hold for any $H\in \mathcal H_{\rm sub}$ if and only if the preference satisfies: 
		(i) ambiguity aversion ($\alpha = 1$) and $u$ is nonincreasing and convex; or (ii) ambiguity proneness ($\alpha = 0$) and $u$ is nondecreasing and concave.
	\end{enumerate}
\end{theorem}
{The corresponding results for supermodular aggregation functions follow by a similar argument; see Appendix~\ref{sec:app_proof}.} We summarize the characterization in the following table.
\begin{table}[h]
	\centering
	\begin{tabular}{c | c | c }
		& supermodular $H$ & submodular $H$ \\
		\hline
		Equity & $\alpha = 0$, or $u$ is nondecreasing and convex & $\alpha = 0$, or $u$ is nonincreasing and convex \\
		\hline
		AAA & $\alpha = 1$, or $u$ is nonincreasing and concave & $\alpha = 1$, or $u$ is nondecreasing and concave
	\end{tabular}
	\caption{Characterization of Equity and AAA in terms of the property of $\Phi_\alpha$}
	\label{table:0}
\end{table}

Theorem \ref{thm:sub} and Table \ref{table:0} show that regardless of the monotonicity of utility, for either submodular or supermodular aggregation functions, equity corresponds to ambiguity or risk proneness, while 
AAA corresponds to ambiguity or risk aversion. A similar message is conveyed in Theorem \ref{thm:ambi} with the sum aggregation function.

To sum up, we come to the   insight that the essential feature of whether equity or AAA  
holds is the ambiguity attitude, while the form of aggregation and the risk attitude also make an impact. 

\section{Conclusion}
\label{sec:conc}

Our paper's main contribution is to  connect equity and ambiguity, two important notions in decision making, which have not been studied together in the literature.
The key message of the paper can be summarized in one sentence: Within the framework of uncertain dependence, equity and aggregate ambiguity avoidance are generally conflicting.
Better equity intrinsically leads to bigger ambiguity in the aggregate loss. 
Specialized to the setting of \cite{K80}, we find that equity is equivalent to ambiguity or risk proneness, while aggregate ambiguity avoidance is equivalent to ambiguity or risk aversion. 
The $\alpha$-maxmin preferences are used to illustrate the above criteria  quantitatively. 
Our analysis considers all dependence structures consistent with the
prescribed marginals together. A possible direction for extension is
to incorporate partial information about dependence and determine
which restrictions preserve aggregation-set inclusion under equalizing
transfers. This would clarify how the equity--ambiguity tradeoff
depends on the information available to the planner.

{\emergencystretch=3em
\subsection*{Acknowledgements}
Y.~Liu acknowledges financial support from the National Natural Science Foundation of China (No.~12401624), The Chinese University of Hong Kong (Shenzhen) University Development Fund (No.~UDF01003336), Guangdong Science and Technology Program (No.~2024QN11X076) and Shenzhen Science and Technology Program (Nos.~RCBS20231211090814028, JCYJ20250604141203005, 2025TC0010), and is partly supported by the Guangdong Provincial Key Laboratory of Mathematical Foundations for Artificial Intelligence (No.~2023B1212010001). 
R.~Wang acknowledges financial support from the Natural Sciences and Engineering Research Council of Canada (CRC-2022-00141, RGPIN-2024-03728).

{

\small

}
}

\appendix
\section{Additional proofs}
\label{sec:app_proof}

\begin{proof}[Proof of Proposition~\ref{prop:1}]
We identify distributions with their associated probability measures. The forward implication in part (i) follows from the expectation characterization of first-order stochastic dominance, and the forward implication in part (ii) follows from set inclusion. 
We prove the two reverse implications by a common argument.
Let $C_b(\R)$ denote the bounded continuous real functions on $\R$, and let $C_b^{\uparrow}(\R)$ denote its subclass of increasing functions. Set $\mathcal L=C_b^{\uparrow}(\R)$ for part (i) and $\mathcal L=C_b(\R)$ for part (ii). In either case, the assumed supremum inequality holds for every $\ell\in\mathcal L$.
Fix $F\in\mathcal M_1$. For each $\ell\in\mathcal L$, define
$$
K_\ell=\{G\in\mathcal M_2:\E_G[\ell]\ge\E_F[\ell]\}.
$$
Since $\ell$ is bounded and continuous, $K_\ell$ is a weakly closed subset of the weakly compact set $\mathcal M_2$. 
{Weak compactness refers to the topology of convergence against bounded continuous functions. In particular, it holds for weakly closed sets of probability laws on a common compact subset of $\R$.} 
We show that these sets have the finite intersection property. Suppose, to the contrary, that $K_{\ell_1}\cap\cdots\cap K_{\ell_r}=\varnothing$ for some $\ell_1,\ldots,\ell_r\in\mathcal L$. The set
$$
C=\left\{\bigl(\E_G[\ell_j]-\E_F[\ell_j]\bigr)_{j=1}^r:G\in\mathcal M_2\right\}
$$
is compact and convex, and is disjoint from the closed convex cone $\R_+^r$. Strict separation in finite dimensions yields a nonzero vector $\mathbf{a} \in \R^r$ such that
$$
\sup_{\mathbf{v} \in C}\mathbf{a}^\top \mathbf{v} < \inf_{\mathbf{w}\in\R_+^r} \mathbf{a}^\top \mathbf{w}.
$$
The right side is finite only if $\mathbf{a} \ge \mathbf{0}$, in which case it equals zero. Therefore $\ell=\sum_{j=1}^r a_j\ell_j$ belongs to $\mathcal L$ and satisfies
$$
\sup_{G\in\mathcal M_2}\E_G[\ell]
<\E_F[\ell]
\le\sup_{F'\in\mathcal M_1}\E_{F'}[\ell],
$$
contradicting the assumed inequality. Thus the finite intersection property holds. Compactness now gives a law $G\in\bigcap_{\ell\in\mathcal L}K_\ell$.

For part (i), this means $\E_G[\ell]\ge\E_F[\ell]$ for every bounded continuous increasing $\ell$, which implies $F\le_{\rm st}G$. Indeed, for each $x\in\R$, the functions $\ell_k(t)=\min\{1,k(t-x)_+\}$ increase pointwise to $\mathbf1_{(x,\infty)}(t)$, yielding $G((x, \infty)) \geq F((x, \infty))$, i.e., $1-G(x) \ge 1-F(x)$. Since $F$ was arbitrary, $\mathcal M_1\precsim\mathcal M_2$.

For part (ii), applying the inequalities to both $\ell$ and $-\ell$ gives $\E_G[\ell]=\E_F[\ell]$ for every $\ell\in C_b(\R)$. Bounded continuous functions determine probability laws on $\R$, so $G=F$. Thus $F\in\mathcal M_2$, and arbitrariness of $F$ proves $\mathcal M_1\subseteq\mathcal M_2$.
\end{proof}

\begin{proof}[Proof of Theorem \ref{thm:all_H}]
	\begin{enumerate}[(i)]
		\item 
		For the ``if" part, note that when the preference is ambiguity prone, we have  $\Phi_0(A) = \sup_{F \in A} \int u \d F$ and by Theorem \ref{th:2}, $\D_H (\boldsymbol{\gamma})\subseteq \D_H (\boldsymbol{\theta})$ whenever $\boldsymbol{\gamma}, \boldsymbol{\theta} \in [0, 1]^n$ satisfying $\boldsymbol{\theta}$ is more equitable than $\boldsymbol{\gamma}$. Therefore, we have  $\Phi_{\alpha}(\D_H (\boldsymbol{\gamma})) \le  \Phi_{\alpha}(\D_H (\boldsymbol{\theta}))$, that is, equity holds.
		
		For the ``only if" part, 
		we show $\alpha=0$ by contradiction. Assume that $\alpha>0$. 
		Since $ u$ is nonconstant,  there exists $a,b \in \R$ such that $u(a)\neq u(b)$.  Without loss of generality, we assume that $u(0)\not= u(1)$ since the general case can be shown by linear transformation (which can be seen at the end of the proof).
		We consider the following cases. Again let $\pbf = (1-\gamma_1, \dots, 1-\gamma_n)$ and $\qbf = (1-\theta_1, \dots, 1-\theta_n)$. 
		\begin{itemize}
			\item [(a)] If $u(0)< u(1)$, we set a concrete setting: $H(x,y)=|x-y|, x,y\in\R$, $\pbf=(0,2p)$ and $\qbf=(p,p)$ with $p\in (0,1/2]$. One can verify that
			$$
			\D_H (\boldsymbol{\gamma}) =\{2p\delta_1+(1-2p)\delta_0\} ~~{\rm and}~~ \D_H (\boldsymbol{\theta}) =\{2s\delta_1+(1-2s)\delta_0,~s\in [0,p] \}.
			$$
			It follows that
			\begin{equation*}
				\begin{aligned}
					\Phi_{\alpha}(\D_H(\boldsymbol{\gamma})) = 2p u(1) +(1-2p)u(0)
				\end{aligned}	
			\end{equation*}
			and
			\begin{equation*}
				\begin{aligned}
					\Phi_{\alpha}(\D_H(\boldsymbol{\theta}))
					& = \alpha \inf_{s\in [0,p]} \{2su(1)+(1-2s)u(0)\} \\
					&~~~+ (1 - \alpha) \sup_{s\in [0,p]} \{2su(1)+(1-2s)u(0)\} \\
					& = 
					\alpha  u(0) +    (1 - \alpha)\{2pu(1)+(1-2p)u(0)\}. 
				\end{aligned}	
			\end{equation*}
			By $u(0)< u(1)$, we have $\Phi_{\alpha}(\D_{H}(\boldsymbol{\theta}))<\Phi_{\alpha}(\D_{H}(\boldsymbol{\gamma}))$  which yields a contradiction to equity.
			\item [(b)] If $u(0)>u(1)$, we set a concrete setting: $H(x,y)=1-|x-y|$, $x,y\in\R$, and $\pbf=(0,2p)$ and $\qbf=(p,p)$ with $p\in (0,1/2]$. One can verify that $\Phi_{\alpha}(\D_{H}(\boldsymbol{\theta}))<\Phi_{\alpha}(\D_{H}(\boldsymbol{\gamma}))$, which yields a contradiction.
		\end{itemize}
		If there exists $a,b$ such that $u(a)\neq u(b)$, then by defining $H(x,y) = a |x-y| +b(1-|x-y|)$ or  $H(x,y) = b |x-y| +a(1-|x-y|)$, we can get  contradiction similarly.  Therefore, we have $\alpha=0$, that is, the preference is ambiguity prone.
		
		\item The result is in parallel with (i).
		
		\item The result is straightforward. \qedhere
	\end{enumerate}
\end{proof}

\begin{proof}[Proof of Theorem \ref{thm:multiply}]
	\begin{enumerate}[(i)]
		\item Denote by $\tilde{x}_1, \dots, \tilde{x}_n$ the Bernoulli random variables with resource profile vector $\boldsymbol{\gamma}$. Again define $\pbf = (p_1, \dots, p_n) = (1-\gamma_1, \dots, 1-\gamma_n)$, whose components indicate the marginal death probability and hence $\tilde{x}_i \sim B_{p_i}$. By direct computation, we have the aggregation random variable
		$$
		H_{\text{a}}(\tilde{x}_1, \dots, \tilde{x}_n) = 1 - \prod_{i = 1}^{n} (1 - \tilde{x}_i) = \left\{
		\begin{aligned}
			& 0, && \text{ if } \tilde{x}_1, \dots, \tilde{x}_n \text{ all take the value } 0;\\
			& 1, && \text{ otherwise}.
		\end{aligned}
		\right.
		$$
		By \eqref{eq:aggregation-new} and \eqref{eq:multiply}, we have 
		$\D_{H_\text{a}}(\boldsymbol{\gamma})=\{\text{cdf of } B_{s}: p_{\text{min}} \le s\le  p_{\text{max}} \}$, where $ p_{\text{min}} =\max\{ p_1, \dots, p_n\}$ and  $ p_{\text{max}} = \min\{1,\sum_{i = 1}^n p_i \}$. 
		We compute:
		\begin{equation}\label{eq:PhiDH}
			\begin{aligned}
				&\quad\quad \Phi_{\alpha}(\D_{H_{\text{a}}}(\boldsymbol{\gamma})) \\
				&= \alpha \inf_{F \in \D_{H_{\text{a}}}(\boldsymbol{\gamma})} \int u \d F + (1 - \alpha) \sup_{F \in \D_{H_{\text{a}}}(\boldsymbol{\gamma})} \int u \d F \\
				&= \alpha \inf_{p_{\text{min}} \le s\le  p_{\text{max}}} \{s u(1) + (1-s) u(0)\}  + (1 - \alpha) \sup_{p_{\text{min}} \le s\le  p_{\text{max}}} \{s u(1) + (1-s) u(0)\}\\
				&= u(0) + \alpha \inf_{p_{\text{min}} \le s\le  p_{\text{max}}} \{s\( u(1) - u(0)\)\} + (1 - \alpha) \sup_{p_{\text{min}} \le s\le  p_{\text{max}}} \{s\left( u(1) - u(0)\right)\} \\
				& = 	\left\{
				\begin{aligned}
					& u(0) +( \alpha  p_{\text{min}} + (1 - \alpha)  p_{\text{max}}) \( u(1)-u(0)\) , &&\text{ if } u(1) \ge  u(0);\\
					&  u(0) +( \alpha  p_{\text{max}} + (1 - \alpha)  p_{\text{min}}) \( u(1) - u(0)\), &&\text{ if } u(1) < u(0).
				\end{aligned}
				\right. 
			\end{aligned}
		\end{equation}
		
		For the ``if" statement, assume that $\boldsymbol{\gamma}, \boldsymbol{\theta} \in [0, 1]^n$ and $\boldsymbol{\theta}$ is more equitable than $\boldsymbol{\gamma}$. Define $\pbf = (p_1, \dots, p_n) = (1-\gamma_1, \dots, 1-\gamma_n)$ and $\qbf = (q_1, \dots, q_n) = (1-\theta_1, \dots, 1-\theta_n)$. Hence, $\qbf$ is more equitable than $\pbf$. Define $q_{\text{min}}$  and $q_{\text{max}} $ for $\qbf$ in parallel with $p_{\text{min}}$  and $p_{\text{max}} $. We know that $p_{\text{min}}\ge  q_{\text{min}}$  and $p_{\text{max}}=q_{\text{max}}$. The result holds trivially for $\alpha=0$. So we consider the case that $\alpha>0$ and $u(1) \le u(0)$. We have $  p_{\text{min}}  \(u(1)-u(0)\)  \le q_{\text{min}}  \(u(1)-u(0)\)$ which implies that $\Phi_{\alpha}(\D_{H_{\text{a}}}(\boldsymbol{\gamma})) \leq \Phi_{\alpha}(\D_{H_{\text{a}}}(\boldsymbol{\theta}))$.

		For the ``only if" statement, it suffices to consider the case $\alpha>0$. Suppose that $\alpha>0$ and $u(1) > u(0)$. Let  $\boldsymbol{\gamma}$ and $\boldsymbol{\theta}$ be such that $\boldsymbol{\theta}$  is more equitable than $\boldsymbol{\gamma}$ and  $p_{\text{min}}> q_{\text{min}}$.  Then we have
		\begin{align*}
			\Phi_{\alpha}(\D_{H_{\text{a}}}(\boldsymbol{\gamma})) & = u(0) +( \alpha  p_{\text{min}} + (1 - \alpha)  p_{\text{max}}) \{ u(1)-u(0)\} \\
			& >  u(0) +( \alpha  q_{\text{min}} + (1 - \alpha)  q_{\text{max}}) \{ u(1)-u(0)\} \\ 
			& = \Phi_{\alpha}(\D_{H_{\text{a}}}(\boldsymbol{\theta}))
		\end{align*}
		which yields a contradiction to equity. 
		Therefore, we have equity holds if and only if $\alpha=0$ or $u(1) \le u(0)$.

		
		\item The proof is similar to that of part (i).
		
		\item The statement is straightforward.\qedhere
	\end{enumerate}
\end{proof}

\begin{proof}[Proof of Theorem \ref{thm:multiply2}]
	The proof is similar to that of Theorem \ref{thm:multiply}. The difference here is that by \eqref{eq:aggregation-new} and \eqref{eq:multiply2}, we have $\D_{H_{\text{b}}}(\boldsymbol{\gamma})=\{B_{s}: p_{\text{min}} \le s\le  p_{\text{max}} \}$, where $ p_{\text{min}} =\max\{\sum_{i = 1}^n p_i - (n-1), 0\}$ and  $ p_{\text{max}} = \min\{p_1, \dots, p_n\}$. 
	
	The other approach is to apply the new aggregation function $x\mapsto 1-H_{\text{a}}(1-x)$ in Theorem \ref{thm:multiply}, where $H_{\text{a}}$ is given by \eqref{eq:multiply}. Recall that $(1-\theta_1, \dots, 1-\theta_n)$ is more equitable than $(1-\gamma_1, \dots, 1-\gamma_n)$ means that $\boldsymbol{\theta}$ is more equitable than $\boldsymbol{\gamma}$.
\end{proof}

\pagebreak[3]
\begin{proof}[Proof of Theorem \ref{thm:sub}]
	\begin{enumerate}[(i)]
		\item It suffices to consider the case that $\alpha>0$.
		For the ``only if" part, first note that both $H_{\mathrm{sum}}(\mathbf{x})=\sum_{i=1}^n x_i$
		and 
        {the extension} $H_{\text{a}}(\mathbf{x})=1-\prod_{i=1}^n (1-
        {[x_i]_0^1})$ are submodular functions, 
        {where $[x]_0^1=\min\{1,\max\{0,x\}\}$; this extension agrees with $H_{\text{a}}$ on Bernoulli inputs}. By Theorem \ref{thm:ambi}, we have $u$ is 
        {discretely} convex. By Theorem \ref{thm:multiply}, we have $u(1) \leq u(0)$. Note that $a H +b$ is submodular whenever $H$ is submodular for any $a>0,b\in\R$. We conclude that $u$ is 
        {midpoint} convex and nonincreasing in $\R$
        {, and hence convex}.
		
		For the ``if'' part, suppose that $u$ is convex and nonincreasing. 
        {We first consider the case $n=2$.} That is, we denote by $H_2=H(1,1)$, $H_1=H(0,1)$ and $H_0=H(0,0)$.
		By the fact that $H$ is submodular, we have $H(1,1) + H(0,0) \leq  H(1,0)+H(0,1)$, that is, $H_2+H_0\le 2H_1$. For the resource profile $\boldsymbol{\gamma}$, we again define $\pbf = (p_1,p_2)  = (1-\gamma_1, 1-\gamma_2)$. We compute the distribution of the aggregate variable ($n= 2$):
		$$
		H(\tilde{x}_1, \tilde{x}_2) = 
		\left\{
		\begin{aligned}
			& H_2, && \text{with probability } s := \p(\tilde{x}_1 = \tilde{x}_2 = 1);\\
			& H_1, && \text{with probability } \p(\tilde{x}_1 = 1, \tilde{x}_2 = 0) + \p(\tilde{x}_1 = 0, \tilde{x}_2 = 1) = p_1 + p_2 - 2 s;\\
			& H_0, && \text{with probability } \p(\tilde{x}_1 = \tilde{x}_2 = 0) = 1 - p_1 - p_2 + s.\\
		\end{aligned}
		\right.
		$$
		Moreover,
		\begin{align*}
			\D_H(\boldsymbol{\gamma}) & = \bigl\{s\delta_{H_2} + (p_1 + p_2 - 2 s) \delta_{H_1} + (1-p_1-p_2+s)\delta_{H_0} :\\
            &\qquad s \in [\max\{p_1+p_2-1,0\}, \min\{p_1, p_2\}] \bigr\}.
		\end{align*}
		Further,
		\begin{align*}
			\Phi_1(\D_H(\boldsymbol{\gamma})) &= \inf_{F \in \D_H(\boldsymbol{\gamma})} \int u \d F \\
			&= \inf_{s \in [\max\{p_1+p_2-1,0\}, \min\{p_1, p_2\}]} \bigl\{ s u(H_2) + (p_1 + p_2 - 2 s) u(H_1)\\
            &\qquad\qquad + (1-p_1-p_2+s) u(H_0) \bigr\}\\ 
			&= \left( (p_1 + p_2)u(H_1) + (1-p_1-p_2)u(H_0) \right) \\
            &\quad \quad + \inf_{s \in [\max\{p_1+p_2-1,0\}, \min\{p_1, p_2\}]} \left\{ s (u(H_2) + u(H_0) - 2 u(H_1)) \right\} 
		\end{align*}
		and
		\begin{align*}
			\Phi_0(\D_H(\boldsymbol{\gamma}))
			&= \left( (p_1 + p_2)u(H_1) + (1-p_1-p_2)u(H_0) \right) \\
            & \quad \quad + \sup_{s \in [\max\{p_1+p_2-1,0\}, \min\{p_1, p_2\}]} \left\{ s (u(H_2) + u(H_0) - 2 u(H_1)) \right\}. 
		\end{align*}
		By the nonincreasing and convexity of $u$ and the submodularity of $H$ ($H_2+H_0\le 2 H_1$), we have
		$$
		u(H_1) \leq u\(\frac{H_2 + H_0}{2}\) \leq \frac{u(H_2) + u(H_0)}{2}. 
		$$
		Hence, 
		$$
		\begin{aligned}
			\Phi_1(\D_H(\boldsymbol{\gamma})) &= \left( (p_1 + p_2)u(H_1) + (1-p_1-p_2)u(H_0) \right)\\
            &\quad + \max\{p_1+p_2-1,0\} \left( (u(H_2) + u(H_0) - 2 u(H_1)) \right);\\
			\Phi_0(\D_H(\boldsymbol{\gamma})) &= \left( (p_1 + p_2)u(H_1) + (1-p_1-p_2)u(H_0) \right)\\
            &\quad + \min\{p_1, p_2\} \left( (u(H_2) + u(H_0) - 2 u(H_1)) \right). 
		\end{aligned}
		$$
		Now we suppose $\boldsymbol{\gamma}, \boldsymbol{\theta} \in [0, 1]^2$ and $\boldsymbol{\theta}$ is more equitable than $\boldsymbol{\gamma}$. Define $\qbf = (q_1, \dots, q_n) = (1-\theta_1, \dots, 1-\theta_n)$ and $\pbf = (p_1, \dots, p_n) = (1-\gamma_1, \dots, 1-\gamma_n)$.  Hence, $p_1 + p_2 = q_1 + q_2$ and $\min\{p_1, p_2\} \leq \min\{q_1, q_2\}$, which implies
		$$
		\Phi_1(\D_H(\boldsymbol{\gamma})) = \Phi_1(\D_H(\boldsymbol{\theta})), \;\; \Phi_0(\D_H(\boldsymbol{\gamma})) \leq \Phi_0(\D_H(\boldsymbol{\theta})).
		$$
		Therefore, equity follows from
		\begin{align*}
			\Phi(\D_H(\boldsymbol{\gamma}))
			= \alpha \Phi_1(\D_H(\boldsymbol{\gamma})) + (1 - \alpha) \Phi_0(\D_H(\boldsymbol{\gamma})) \le\Phi(\D_H(\boldsymbol{\theta})).
		\end{align*}
        {For general $n$, symmetry gives $H(\tilde{x}_1,\ldots,\tilde{x}_n)=h(\sum_i\tilde{x}_i)$, where $h(k)$ is the value of $H$ at a binary vector with $k$ ones. Submodularity yields $h(k)+h(k+2)\le2h(k+1)$ for $k=0,\ldots,n-2$. The same convexity inequality as above shows that $v(k)=u(h(k))$ is discretely convex. The sufficient-part argument in the proof of Theorem~\ref{thm:ambi}(i), applied to $v$ and sum aggregation, therefore proves equity for arbitrary $n$.}

        \item The proof is similar to that of part (i). 
        {For sufficiency, when $u$ is nondecreasing and concave, $v=u\circ h$ is discretely concave. By the extremal-expectation argument in the proof of Theorem~\ref{thm:ambi}, its best-case expectation depends only on $\sum_i p_i$. Thus, for equal-total profiles with nested $H$-aggregation sets, the best-case value is unchanged and the worst-case value cannot increase, proving AAA.}

		\item The statement is straightforward.
	\end{enumerate}
{For the supermodular column of Table~\ref{table:0}, $h$ is nondecreasing and discretely convex. Nondecreasing convex $u$ makes $u\circ h$ discretely convex, while nonincreasing concave $u$ makes it discretely concave; the same extremal-expectation argument proves sufficiency. Necessity follows from $H(\mathbf x)=a+(b-a)(\sum_i[x_i]_0^1-n+1)_+$ for $a<b$, which forces the indicated monotonicity, and $H=a+d\sum_i x_i$ for $a\in\R$ and $d>0$, which forces the indicated curvature.}
\end{proof}

\end{document}